\documentclass[preprint,12pt]{elsarticle}

\usepackage{filecontents}

\usepackage{mathtools}
\usepackage{siunitx}
\usepackage{subcaption}
\usepackage{tikz}
\usepackage{overpic}
\usepackage{import}
\usepackage{here}
\usepackage{pgf}
\usepackage[utf8]{inputenc}
\usepackage{svg}
\usepackage{bibspacing}

\usepackage{booktabs}
\usepackage{enumitem}
\usepackage{geometry}
\usepackage{lineno}
\usepackage{svg}
\usepackage{algorithm}
\usepackage{algpseudocode}
\usepackage[version=3]{mhchem}
\usepackage{amsmath,amsthm,amssymb,amsfonts,amsbsy,txfonts}
\usepackage{mathrsfs}
\usepackage{comment}
\usetikzlibrary{matrix}

\usepackage{multicol}
\usepackage{nomencl}
\usepackage{etoolbox}

\DeclarePairedDelimiterXPP\BigOSI[2]%
  {\mathcal{O}}{(}{)}{}%
  {\SI{#1}{#2}}
\usepackage{hyperref}

\usepackage{amssymb}

\tikzset{ 
    table/.style={
        matrix of nodes,
        row sep=-\pgflinewidth,
        column sep=-\pgflinewidth,
        nodes={
            rectangle,
            draw=black,
            align=center
        },
        minimum height=1.em,
        text depth=0.5ex,
        text height=1ex,
        nodes in empty cells,

        row 1/.style={
            nodes={
                fill=black,
                text=white,
                font=\bfseries
            }
        }
    }
}

\journal{Combustion and Flame}

\begin{document}

\begin{frontmatter}

\title{
Assessment of dynamic adaptive chemistry with tabulated reactions for the simulation of unsteady multiregime combustion phenomena}

\author[label1]{A. Surapaneni}
\author[label1]{D. Mira\corref{cor1}}
\cortext[cor1]{Corresponding author}
\ead{daniel.mira@bsc.es}

\address[label1]{Barcelona Supercomputing Center (BSC),
Plaça Eusebi Güell, 1-3
08034, Barcelona, Spain}

\date{July 2022}

\journal{Combustion and Flame}
\address{}

\begin{abstract}

Solving chemistry is an integral part of reacting flow simulations, usually dominating the computational cost. Among the different strategies to accelerate the solution of chemistry and to achieve realizable simulations, the use of Dynamic Adaptive Chemistry (DAC) stands out among other methods. DAC methods are based on the use of reduced mechanisms generated from local conditions. The reduction process is computationally expensive and strategies for reducing the frequency of reduction and the re-utilization of the generated reduced mechanisms are key in making DAC methods computationally affordable. In this study, a new method hereby referred as Tabulated Reactions for Adaptive Chemistry (TRAC) is proposed to correlate chemical states with their reduced mechanisms in order to reduce both the frequency of reduction and to allow for re-utilization of reduced chemical schemes. TRAC introduces a mechanism tabulation strategy based on the use of a low-dimensional space that defines the thermo-chemical conditions for which specific reduced reaction mechanisms are stored.
Chemistry reduction is achieved by the use of Path Flux Analysis (PFA) with a reaction rate-sensitivity method to achieve further reduction in the reaction mechanisms. The new TRAC proposal is applied to various canonical transient problems and the results are compared with reference solutions obtained from detailed chemistry calculations. A speedup of about 4x was achieved with TRAC while maintaining an error under $3\%$ in the prediction of the major and minor species, flame structure, and flame propagation.

\end{abstract}


\end{frontmatter}


\section{Introduction}
\label{}
The dynamics of chemically reacting flows are determined by interactions between thermo-chemical states with the flow field. These interactions can lead to unsteady effects like ignition, extinction, and complex phenomena like partially premixed flame propagation or multi-mode combustion ~\cite{lu2009toward}. Chemistry can play a key role in such conditions, so a detailed description of the chemical processes is important to accurately predict these phenomena. However, including detailed chemistry in numerical simulations still remains
computationally expensive, especially when dealing with complex fuels and turbulent flow conditions~\cite{MIRA2022}. It is for such conditions, that methods for reducing the cost of solving the chemistry are of high importance. One possibility to reduce this cost is to identify and neglect the species of lower relative importance while solving the chemistry. This approach also referred to as Dynamic Adaptive Chemistry (DAC), is based on the use of reduced reaction mechanisms generated from local conditions to describe the combustion chemistry. The other possibility to mitigate the computational costs is to reduce the dimensionality of the problem by clustering~\cite{wang2019species} or tabulation~\cite{lu2009toward,pope2009efficient,Oijen2016}.

Cell clustering can be used to accelerate chemistry integration by grouping cells with similar thermo-chemical states. The extracted data from the clusters is then transformed into thermo-chemical reactors, which are then solved using ordinary-differential equations (ODE) solvers. The clustering can be based on graph partition methods \cite{wang2019species}, Principal Component Analysis (PCA) \cite{d2020adaptive}, or by a bounded k-means algorithm \cite{perini2013high}. After the chemistry is solved in the lower-dimensional cluster space, the data obtained is re-interpolated back to the original computational grid. Clustering methods have shown significant speedups ranging from 4x to 20x, based on the level of required accuracy and complexity of the chemical problem. Methods based on clustering were applied to homogeneous auto-ignition problems \cite{wang2019species,d2020adaptive}, Large Eddy Simulations (LES) of laboratory flames \cite{franke2017tabulation} and Direct Numerical Simulations (DNS) \cite{chi2021fly}. Clustering methods rely entirely on the definition of the clusters and the number of groups. This aspect requires the identification of correlation between states, which is often achieved by PCA \cite{d2020adaptive} or more recently by Artificial Neural Networks (ANN's) \cite{franke2017tabulation,galassi2022adaptive,nguyen2021machine}. These methods are robust but require a certain amount of user intervention to ensure the optimal definition of the clustering. Moreover, as the demand for accuracy increases, the speedup gained by such methods reduces. In fact, in multiregime conditions, the definition of the clusters based on thermo-chemical composition can be cumbersome as multiple correlated chemical states can react differently based on the mixture fraction gradient they perceive.

Unlike the clustering method, DAC methods aim to reduce the computational cost by solving for a reduced system of unknowns locally. To achieve this, an ideal DAC method would perform a chemistry reduction step at every computational point during runtime, so the resulting reduced reaction mechanisms are representative of the local conditions. The reduction process is usually based on methods that truncate reaction pathways like Directed Relation Graph (DRG) \cite{lu2005directed}, Directed Relation Graph Error Propagation (DRGEP) \cite{pepiot2008efficient} or Path Flux Analysis (PFA) \cite{sun2010path}. However, computing reduced mechanisms at every grid point and at every time step becomes unfeasible due to high computational costs. As the reduction is based on the identification of relevant species and their derived reactions, the reaction mechanism at each computational cell only needs to be updated after substantial evolution in the local thermo-chemical state. An error function can then be used to control the frequency of the reduction so important savings in computational time can be achieved. This holds true provided that a good temporal correlation function can be defined \cite{sun2015multi}. This idea can also be extended to identify spatial correlations between similar chemical states which introduces the concept of correlated DAC or CODAC methods \cite{sun2015multi}. A limiting case of this family is the tabulated DAC method, also referred to as TDAC, where the reaction mechanisms are tabulated using error functions, and no reduction strategy is needed after a certain time. 

Tabulation methods with DAC were first described by Contino et al.~\cite{contino2011coupling} and aimed to couple the In Situ Adaptive Tabulation (ISAT) method with DAC to dynamically identify and store unique chemical states. These states are then retrieved or updated accordingly as the simulation proceeds. Contino et al. tested the TDAC method in a Homogeneous Charge Compression Ignition case (HCCI) and reported a speed-up of around 300x. Later, Contino et al. extended the TDAC method to include reduced chemistry using the DRG method \cite{contino2012simulations}, which showed a speedup of 500x in HCCI simulations of diesel. The TDAC method has also been successfully applied to other flow conditions such as in supersonic combustion or moderate/intense low-oxygen dilution combustion \cite{li2018assessment,wu2018application}. These TDAC methods were based on the tabulation, retrieval, and recycling of chemical states. Such methods are bound to run into efficiency issues as the number of unique and uncorrelated states increase, which are often encountered in multiregime problems. From this perspective, the use of a DAC-based strategy for the tabulation of reduced chemical schemes can provide a more general framework to deal with unsteady multiregime combustion phenomena.

DAC-based approaches have shown high potential to be used in DNS and LES of turbulent flames \cite{gou2013dynamic,sun2017multi,yang2013dynamic,yang2017parallel}. The flow in such conditions is inherently unsteady, as the reacting layer interacts with the constantly changing turbulent flow field. The combustion process can move from one regime to another, and unsteady processes can occur depending on the local conditions. In order to avoid the aforementioned problems and still achieve a high level of reduction, a new approach based on the Tabulation of Reactions for Adaptive Chemistry (TRAC) is proposed. This method is developed to reproduce unsteady multiregime combustion phenomena without requiring on-the-fly chemistry reduction. Details of this method and its integration with PFA is discussed in the following sections.

The paper is structured as follows. First, the chemical reduction strategy employed in the proposed DAC framework is presented and discussed. It is followed by the description of the TRAC proposal along with a classical method based on temporal correlation functions that is used for validation purposes. A posterior validation of the TRAC approach is conducted for benchmark problems. Solutions obtained by TRAC are compared with the results from detailed chemistry and CODAC using the same flow solver. To conclude, an analysis of the computational cost is given with final remarks and directions for future work.

\section{Chemistry reduction strategy}\label{subSec_PFA}\addvspace{10pt}

Despite the fact that various methodologies for chemistry reduction have been developed, methods based on truncating chemical pathways like DRG \cite{lu2005directed}, DRGEP \cite{pepiot2008efficient} and PFA \cite{sun2010path} are usually favored in DAC applications due to their simple integration with on-the-fly chemistry reduction techniques. PFA is based on computing a relation matrix that numerically quantifies the relation amongst species at any given thermo-chemical state. This matrix is built from a predefined set of key species and species having an influence on the key species through intermediate pathways. This results in a high-order relation matrix with the inclusion of multiple generations of reaction pathways~\cite{sun2010path}. In this study, PFA is proposed as the chemistry reduction method due to the possibility of inclusion of intermediate reacting paths and the extension to higher-order relations amongst species. In PFA reaction pathways are classified based on their relative importance to a set of pre-defined key species. A threshold is then used to discriminate the most relevant reactions/species that can be used to describe the local chemical evolution of the given state. The chemistry reduction strategy proposed in this study is based on the use of an optimized implementation of PFA~\cite{sun2010path} combined with Reaction Flux Analysis (RFA)~\cite{rfabook}.

In PFA, a set of key species is used to identify the most relevant reaction paths and long reaction chains. The reference mechanism is reduced for a given thermodynamic state, so a chemical scheme with most relevant species and reactions is obtained. PFA includes both consumption and production pathways, which allows for controlling catalytic effects over multiple generations~\cite{sun2010path}.

The production $P_{k}$ and consumption rates $C_{k}$ of species $k$ through $N_{r}$ reactions is given by: 

\begin{equation}
P_k = \Sigma_{r=1}^{N_r} \max(0 ,v_{k,r} \mathcal{Q}_r),
\label{eq:pa}
\end{equation}

\begin{equation}
C_k = \Sigma_{r=1}^{N_r} \max(0 ,-v_{k,r} \mathcal{Q}_r).
\label{eq:ca}
\end{equation}

where $r$ is the index of the reaction in the mechanism, $N_{r}$ is the total number of elementary reactions in the mechanism, $v_{k,r}$ is the stoichiometric coefficient of species $k$ in reaction $r$, and $\mathcal{Q}_r$ is the net rate of progress of reaction $r$. 
 
Production $P_{kj}$ and consumption $C_{kj}$ of species $k$ in relation to species $j$ is given as:

\begin{equation}
P_{kj} = \Sigma_{r=1}^{N_r} \max(0 ,v_{k,r} \mathcal{Q}_r  \delta_{j}^{r}),
\label{eq:pab}
\end{equation}

\begin{equation}
C_{kj} = \Sigma_{r=1}^{N_r} \max(0 ,-v_{k,r} \mathcal{Q}_r \delta_{j}^{r}).
\label{eq:cab}
\end{equation}

Here the Dirac delta $\delta_{j}^{r}$ indicates the involvement of species $j$ in reaction $r$. The resulting relation matrix is then normalized by the absolute maximum between the production and consumption rates of species $k$, resulting in the $1^{st}$ generation of normalized relation coefficients $\lambda^{p-1st}_{kj}$ and $\lambda^{c-1st}_{kj}$. Those are given by:

\begin{equation}
\lambda^{p-1st}_{kj} = \frac{P_{kj}}{\max(P_{k},C_{k})},
\label{eq:rp1}
\end{equation}

\begin{equation}
\lambda^{c-1st}_{kj} = \frac{C_{kj}}{\max(P_{k},C_{k})}.
\label{eq:rc1}
\end{equation}

PFA allows for extension to include second-generation species, which can have an impact on the key species through intermediate arbitrary species $i$. The second generation coefficients $\lambda^{p-2nd}_{kj}$ and $\lambda^{c-2nd}_{kj}$, which take into account all possible reactions pathways between species $k$ and $j$ through intermediate species $i$ are given by:

\begin{equation}
\lambda^{p - 2nd}_{kj} = \Sigma_{i \neq k.j} (\lambda^{p-1st}_{ki} \lambda^{p-1st}_{ij}),
\label{eq:rp2}
\end{equation}

\begin{equation}
\lambda^{c - 2nd}_{kj} = \Sigma_{i \neq k.j} (\lambda^{c-1st}_{ki} \lambda^{c-1st}_{ij}).	
\label{eq:rc2}
\end{equation}

This step can be extended to any arbitrary number of generations. The overall relation matrix $\Lambda_{kj}$ is finally constructed by summing up all the reaction coefficients as follows:

\begin{equation}
\Lambda_{kj} = \lambda^{p-1st}_{kj} + \lambda^{c-1st}_{kj} +  \lambda^{p-2nd}_{kj} + \lambda^{c-2nd}_{kj}.
\label{eq:rab}
\end{equation}

After computing the relation matrix ($\Lambda_{kj}$), various algorithms can be applied to eliminate species. These include Depth First Search (DFS), Breadth First Search (BFS), and the Dijkstra algorithm, among others \cite{niemeyer2011importance}. Unlike PFA, the algorithm for removal demands more importance in graph-based chemistry reduction methods like the DRGEP, as is explored in \cite{niemeyer2011importance}. In the current study a species $k$ and its relevant reactions are eliminated if the maximum of all its relation coefficients with respect to all the key species falls below the PFA threshold ($.pdfilon_{PFA}$). A species $k$ is removed from the reaction mechanism if the following expression holds:

\begin{equation}
.pdfilon_{PFA}  > \max(\Lambda_{kj}),
\label{eq:nax}
\end{equation}

where $j \in S_{key}$ and $S_{key}$ is the predefined set of key species.

As PFA is based on local thermo-chemical states, it is difficult to define a universal threshold that could control the error propagation during the reduction process as it occurs with DRGEP. However, it is observed 
that values obtained from the relation matrix ($\Lambda_{kj}$) are better resolved in a log scale rather than a linear scale. Furthermore, it was noticed that when the PFA threshold ($.pdfilon_{PFA}$) was varied in exponential s.pdf, the error of reduction correlated better with the threshold. The relation matrix is also normalized by its maximum to account for variability in $\Lambda_{kj}$ among different thermo-chemical states and the PFA threshold is modified as:

\begin{eqnarray}
{\log({.pdfilon^*_{PFA}})} = {-h(1-.pdfilon_{PFA})},
\label{eq:rablog}
\end{eqnarray}

where ${.pdfilon^*_{PFA}}$ is the modified threshold, $.pdfilon_{PFA}$ is the usual PFA threshold as specified in \cite{sun2010path} and $h$ is constant set to 10 in this study. This modification of the threshold accounts for the exponential nature of the relation matrix and makes sure that the threshold for PFA always lies between 0 and 1. We elude to this modification later in the section, where we justify this choice by presenting results from a homogeneous auto-ignition problem at various levels of PFA reduction.

The PFA reduction can be further extended by the use of Reaction Flux Analysis (RFA)~\cite{rfabook}. RFA refers to a chemistry reduction technique based on 
the overall contribution of certain reaction rates to the production or consumption of the species of interest. This reduction strategy is well suited to be combined with PFA as it permits to further reduce the chemistry of the problem.

In RFA, reactions are ordered by their respective contributions to each of the species, so a threshold can be used to discriminate the most relevant reactions. The RFA method is quite powerful in reducing the number of reactions, but it can lead to substantial errors. The error is controlled by using the RFA method dynamically where local thermo-chemical states are updated during the reduction process. Normalised contribution to the production ($\gamma_{k,r}^+$) and consumption ($\gamma_{k,r}^-$) of species $k$ in reaction $r$ is computed as:

\begin{equation}
  \begin{array}{l}
 \gamma_{k,r}^+ = v_{k,r} \mathcal{Q}_r / {P_k}, \\
 \gamma_{k,r}^- = -v_{k,r} \mathcal{Q}_r / {C_k}.
  \end{array}
\label{eq:rfa}
\end{equation}

 If the maximum of these contributions falls below the RFA threshold ($.pdfilon_{RFA}$), reaction $r$ is removed from the mechanism. RFA and PFA have common operations and when used together, can lead to significant reduction in the CPU cost due to the recycling of variables and operations. In order to take advantage of the common operations, the computation of consumption and production rates of species $k$ is plugged into the RFA algorithm. A speedup in the order of ($H$$ \, N_{sp}$) can be achieved where $N_{sp}$ corresponds to the total number of species ($k=1, \cdots, N_{sp}$) and $H$ is the average number of reactions in which each species participates. For modern detailed mechanisms, $H$ is around 5 \cite{lu2009toward}.

\begin{figure} [H]
  \centering
  \includegraphics[width=0.5\textwidth]{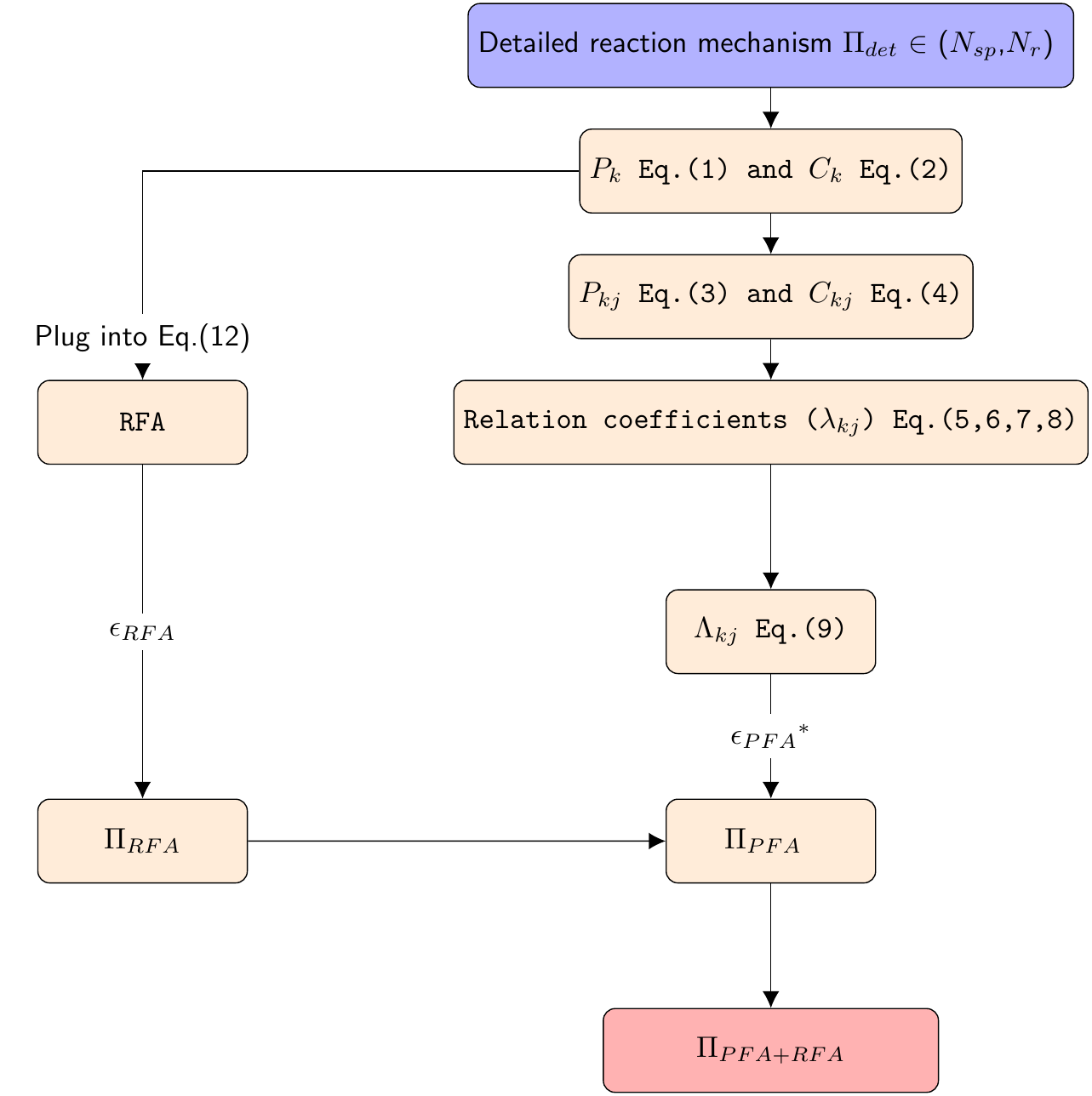}
  \caption{Combined PFA-RFA algorithm, previously defined notation apply with $\Pi_{det}$, $\Pi_{PFA}$, $\Pi_{RFA}$, $\Pi_{PFA+RFA}$ additionally representing chemical schemes respectively.}
  \label{fig:pfarfa}
\end{figure}

The algorithm of the combined chemistry reduction strategy using both PFA and RFA is shown in Fig.~(\ref{fig:pfarfa}). The variables $P_{k}$ and $C_{k}$, which represent the overall production and consumption of species $k$ respectively, are used in both algorithms and are hence computed only once when PFA is combined with RFA.

\begin {figure}[H]
\includegraphics[width=1.0\textwidth]{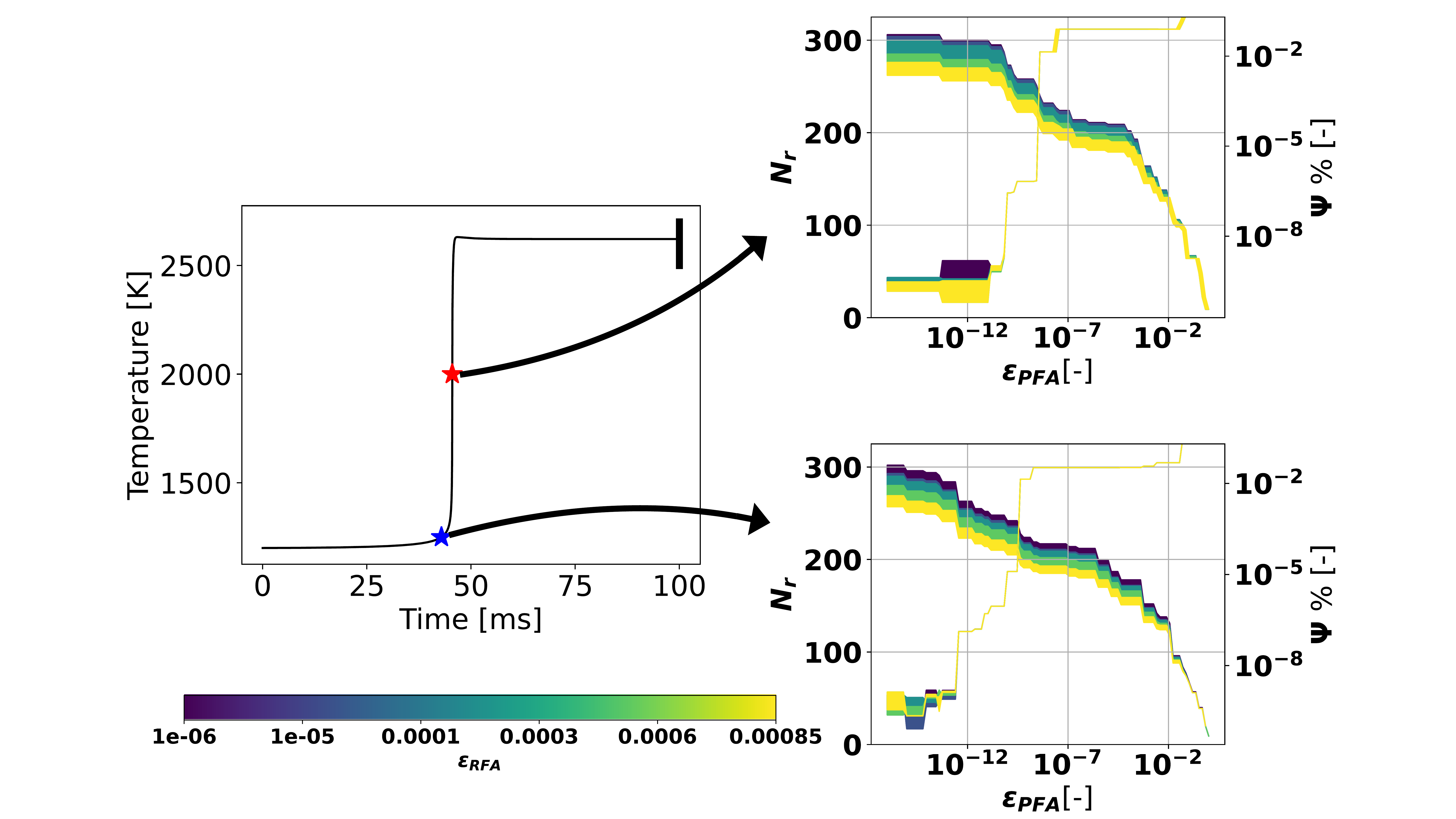}
\caption{Temporal evolution of mixture temperature in an auto-ignition problem (left). Number of reactions $N_{r}$ and error $\Psi$ computed by Eq.~(\ref{eq:error2}) at various PFA and RFA thresholds applied to pre-ignition (top right) and ignition states (bottom right), respectively for a temporally evolving homogeneous reactor.}
\label{fig:hrrpfa}
\end {figure}

The definition of the PFA and RFA thresholds are critical aspects of the method and need to be well defined to achieve the maximum possible reduction with error control. The influence of these thresholds on the resulting number of species and associated error is shown in Fig.~(\ref{fig:hrrpfa}). The error due to reduction is defined as the maximum deviation of any state variable compared to the detailed chemistry solution. The error is mathematically defined as: 
\begin{equation}
    \Psi = \max \left(\frac{\psi_{Ref}- \psi_{Red}}{\psi_{Ref}}\right),
    \label{eq:error2}
\end{equation}

where $\Psi$ is the error, $\psi_{Ref}$ and $\psi_{Red}$ correspond to state variables, namely, temperature and species mass fractions, obtained in the reference and the reduced solutions respectively. 

Here, the present reduction strategy is applied to two thermo-chemical states during the transient evolution of an auto-igniting homogeneous reactor. The first state is the pre-ignition state, which is defined at a time $t$ when the temperature of the homogeneous reactor increases by $100$ K, starting from the initial temperature. The second state is during ignition and is defined at time $t = t_{ign}$, which corresponds to the time at the highest rate of temperature rise.

In Fig.~(\ref{fig:hrrpfa}) it can be seen that exponential s.pdf are taken in the PFA threshold, this was motivated by the fact that values of the entries in the relation matrix ($\Lambda_{kj}$) were found to be better represented in log scale than in linear scale. Moreover, it was found that when exponential s.pdf were taken, changes in the error of reduction were better captured. This prompted the modification of the PFA threshold as proposed in Eq. (\ref{eq:rablog}). Furthermore, it can also be noted that there always exists a value of the PFA threshold at which the reduction error increases substantially, the log-log nature of this relation indicates the proximity among these points. Hence, a global PFA threshold can be defined wherein the error introduced by the reduction can be controlled. The error introduced at various RFA thresholds is almost negligible, mainly due to the fact the RFA reduction retains reaction pathways involving key species and hence, limiting the error. This behavior is evidenced by the reduced intensity of reduction by RFA at higher PFA thresholds, where most of the retained reactions involve the key species. The homogeneous reactor tests were extended to different lean and rich compositions using different chemical mechanisms to identify a possible lack of correlation. The results did not show fundamental differences compared to the case included in the paper, so these figures were omitted for the sake of brevity. The reader is referred to the supplementary material for additional results. 

\section{Theoretical framework for dynamic adaptive chemistry}\label{Sec_DAC}\addvspace{10pt}

The mass balance in combustion problems is described by the conservation equation for mass factions of chemical species $Y_k$, shown in Eq. (\ref{eq:Yk}):

\begin{eqnarray}
\label{eq:Yk_trans}
\frac{\partial (\rho Y_k)}{\partial t} + \boldsymbol{\nabla} \cdot \left( \rho 
\boldsymbol{u} Y_k \right) = -\boldsymbol{\nabla}  \cdot \left( \rho 
\boldsymbol{V_{k}} Y_k \right) + \dot{\omega}_{k},
\label{eq:Yk}
\end{eqnarray}

where $\rho, t, \boldsymbol{u}$ and $\dot{\omega}_{k}$ are given in standard notation and represent the density, time, velocity vector, and chemical source term respectively. The first term on the right-hand side describes the diffusion of species given by the diffusion velocity $\boldsymbol{V_{k}}$, often described by Fick’s law~\cite{lu2009toward}. The solution of this equation involves the evaluation of the transport processes (convection and diffusion), and the chemical source integration. Stiffness introduced by the chemical source term prevents the use of explicit methods, hence a first-order splitting scheme \cite{pope2009efficient} is used to separate the transport from chemical integration and solve the stiff  chemical source implicitly. Integration of the chemical source term is known to scale by $\mathcal{O}$$({N_{sp}}^3$), where $N_{sp}$ is the number of species in the reaction mechanism. This cubic dependency can be mitigated by the use of analytical jacobians, where the order reduces to $\mathcal{O}$(${N_{sp}}^a$), where $a$ takes values between 2 and 3 ~\cite{lu2009toward}. DAC methods aim to reduce this cost by using a locally generated reduced mechanism for the chemical source integration.

The automatic generation of first-order (first generation) reduced mechanisms requires [$N_{key}$ $\times$ $N_{sp}$] evaluations to compute the relation coefficients, where $N_{key}$ is the number of key species and $N_{sp}$ is the total number of species in the reaction mechanism. For higher-order coefficients, the number of operations scales as $2^{G}$, where $G$ is the number of generations. The computation of the relation matrix can be very expensive as the reference mechanism size increases, so reducing the frequency of the reduction process is essential to achieve high computational efficiency. The limit of this approach is the complete avoidance of the chemistry reduction step and the use of pre-defined reaction mechanisms that can be calculated and stored in a pre-processing stage. 
This limit is explored in mechanism tabulation approaches~\cite{an2019dynamic, zhou2016chemistry}, which are very attractive for chemistry adaptation. Most of the tabulated approaches require a preliminary analysis of the chemistry and identification of the relevant species to define appropriate error functions to control the tabulation. This identification can get complex when dealing with realistic fuels and usually requires \emph{a priori} knowledge of the relevant species and radicals involved in certain phenomena. 

The following sections present two methods of using DAC in reacting flow simulations. First, a widespread DAC method based on correlation functions, also known as CODAC, and second, a new strategy for retrieving locally-reduced chemical reaction mechanisms based on a low-dimensional thermo-chemical database, hereby referred as TRAC. 

\subsection{Correlated dynamic adaptive chemistry (CODAC)}\label{Sec_DAC}\addvspace{10pt}

In this study, the reference DAC method is based on the assumption that similar reduced schemes can be applied to states, which are thermo-chemically correlated.
This correlation can be either in space and/or in time and based on a set of species and mixture temperature that are used to define a correlation function, as proposed by Sun et al.~\cite{sun2015multi}. The species are chosen in such a way that they accurately represent the chemistry of the state, Sun et al.~\cite{sun2015multi} proposed, mass fractions of the fuel, $\mathrm{CH_{2}O}$, $\mathrm{OH}$ and $\mathrm{HO_{2}}$ to describe the thermo-chemical correlation. In this study, the same correlation function as in ~\cite{sun2015multi} is retained, but with the addition of $\mathrm{CO}$ to account for rich chemistry and post-ignition kinetics. The correlation function is defined as :

\begin{equation}
\begin{aligned}
\Delta = max \left (\frac{T_{n+1}-T_{n}}{\zeta}, 
\frac{ln Y_{k_{n+1}}-ln Y_{k_{n}}}{\zeta} \right ),
\end{aligned}
\label{eq:codacnew}
\end{equation}  

where $\Delta$ is the correlation function, T is the temperature, n is the time instance, $\zeta$ is the user-specified correlation threshold, and $Y_{k}$ are the species involved in the correlation function with $k$ = $\mathrm{CH_{2}O}$, $\mathrm{OH}$, $\mathrm{CH_{4}}$, $\mathrm{HO_{2}}$ and $\mathrm{CO}$. States are identified as correlated if $\Delta$ $<$ 1 and the reduced mechanism from the previous reduction is retained. The reduction process in the CODAC method then relies entirely on the definition of the correlation function, which needs to be redefined for every unique problem. Furthermore, to capture rapid changes in chemistry during transient events like auto-ignition or extinction, the threshold for  the error function needs to be lowered further increasing the number of reduction subs.pdf, making the method more computationally expensive \cite{liu2018modelling}.

\subsection{Tabulated Reactions for Adaptive Chemistry  (TRAC).}\addvspace{10pt}

This section introduces a new strategy for performing DAC calculations without the need for on-the-fly chemistry reduction, which can be computationally expensive for large reaction mechanisms. TRAC proposes an algorithm for storing locally-reduced chemical reaction mechanisms as an alternative to the use of error functions~\cite{an2019dynamic}. The concept of low-dimensional manifolds is used to relate multi-variable data to a few control variables to identify regions with different chemical activities. Tabulation in this regard is more global and hence, can be generated in a pre-processing stage. Using this generalization, a wide range of conditions can be represented provided that appropriate controlling variables are selected to represent the multi-dimensional space. 

A conventional approach to describe the representative thermo-chemical states in non-premixed combustion is the use of the mixture fraction $Y_{\xi}$ and a chemical progress variable $Y_{C}$, which have been successfully applied in flamelet methods~\cite{fiorina2003modelling,illana2021extended,van2001modeling}. This approach can be retained here to define a two-dimensional thermo-chemical database that can be discretized along these two quantities. Reduced schemes can be generated for each entry of the database and stored on the local memory of the node. TRAC uses this thermo-chemical database to store and retrieve the reduced reaction mechanisms, which are constructed either on-the-fly or \emph{a priori} using homogeneous reactors. In case the table is generated on-the-fly, a few iterations are usually sufficient to generate reduced reaction mechanisms for majority of the states in the thermo-chemical space. The locally reduced reaction mechanisms can then be accessed by the mixture fraction and progress variable. 

\begin {figure}[H]
\centering
\includegraphics[width=.65\textwidth]{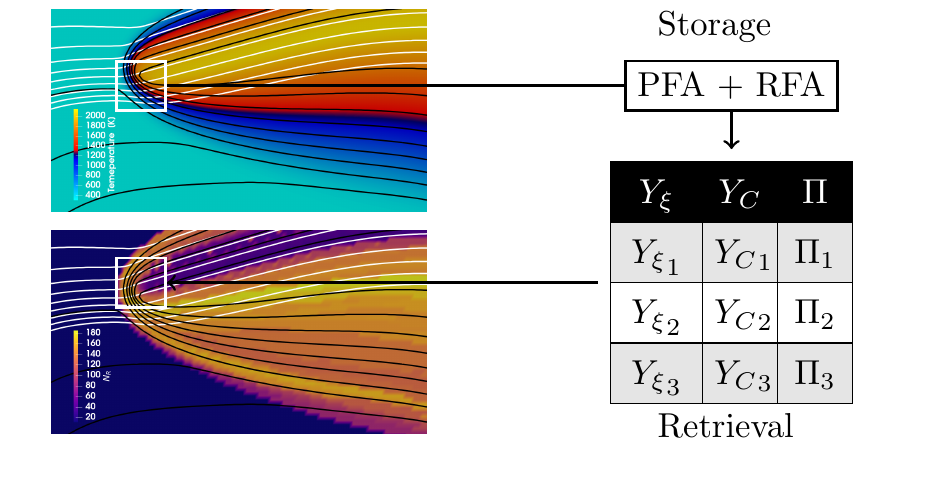}
\caption{TRAC representation, temperature (top), number of reactions $N_{r}$ (bottom) in a triple flame. White and black lines represent contours of iso $Y_{\xi}$ and $Y_C$, respectively. $\Pi$ is the tabulated reduced chemical scheme.}
\label{fig:tadcVV}
\end {figure}

In the TRAC method, the flame structure is not prescribed, as it occurs in flamelet methods, where a pre-defined flamelet configuration, usually premixed or non-premixed \cite{illana2021extended,van2001modeling,vreman2008premixed} is used to tabulate the chemical source terms and transport properties. Unlike traditional flamelet methods \cite{fiorina2003modelling,illana2021extended,van2002flamelet}, which often require additional modeling to recover the correct thermo-chemistry in multiregime conditions, the TRAC method computes these thermo-chemical states using local conditions. The chemistry is resolved by integrating the chemical source terms obtained through the tabulated reaction mechanisms without the need for on-the-fly reduction. Since the tabulation is for reaction mechanisms, TRAC tables have lower memory footprints than ISAT or flamelet tables. Note that as the transport is solved with the reference mechanism, the effects of detailed transport (diffusion and advection) are taken into account. The TRAC method can be visualized in Fig.~(\ref{fig:tadcVV}) using a triple flame as an example.

In summary, two strategies of DAC methods to be used in detailed chemistry simulations are considered. It includes the new proposal for TRAC, which tabulates the reaction mechanisms based on a set controlling variables and stores them into a low-dimensional database, and the CODAC method which is based on the temporal correlation of thermo-chemical states. The reduction strategy is the same for both methods, though the key difference is the frequency of the reduction strategy and the use of a correlation function.

\section{Numerical setup}\label{Sec:3}\addvspace{10pt}

The following section is dedicated to describing the numerical approach used to solve the computational problems. It includes the description of the flow and the chemistry solvers and provides specific details of both the PFA and TRAC methods.
\subsection{Flow and chemistry solvers}

The computations in this study are carried out using the multi-physics code Alya \cite{vazquez2016alya}, which is based on the low-Mach number approximation of the Navier-Stokes equations. Additional conservation equations are solved for enthalpy and chemical species. Chemistry is decoupled from the transport using operator splitting~\cite{ren2014use} techniques to allow the use explicit schemes. The transport is solved under the unity Lewis number assumption. The system of equations is discretized using Finite Elements (FE) with low-dissipation numerical schemes for low-Mach and scalar transport~\cite{benajes2022analysis,both2020low,mira2020numerical}. A third-order Runge-Kutta is used for temporal integration for momentum and scalar transport \cite{both2020low}. For chemical integration and calculation of the source terms, the implicit solver CVODE \cite{cohen1996cvode} and the semi-implicit solver  ODEPIM \cite{yang2017parallel} are couple with Alya.

 \subsection{PFA reduction}

  For the conditions of this study, a $1^{st}$-order PFA reduction yielded an accurate prediction of the burning rates for the relevant species, while achieving high levels of chemistry reduction at a low computational cost. Therefore, a single generation of relations amongst the species is considered for all the computational cases.

 The key species of the PFA algorithm are defined considering the major species encountered in typical methane-air flames: $\mathrm{CH_{4}}$, $\mathrm{O_{2}}$, $\mathrm{CO_{2}}$, $\mathrm{H_{2}O}$ and $\mathrm{CO}$. Even though, optimized identification of the key species will yield a higher degree of reduction, the definition is kept simple to illustrate the robustness of the PFA method, which can classify intermediate reaction pathways just based on major chemical pathways leading from reactants to products. The threshold for the reduction methods namely, PFA and RFA were set to $.pdfilon^*_{PFA}=0.85$ and $.pdfilon_{RFA}=0.01$, respectively. Note that the value used for {$.pdfilon^*_{PFA}$} means a truncation of the reaction pathways with normalized reactivities less than 85 $\%$ of the most reactive pathway. The RFA threshold directly corresponds to an error of $1$ $\%$.
 
 After the chemistry reduction step, a set of retained $Y_{k}^{Ret}$ and removed species $Y_{k}^{Rem}$ is obtained. During chemical integration, $Y_{k}^{Rem}$ are treated as \emph{frozen} in chemistry by forcing their chemical source terms to 0. A similar approach is used in other DAC methods \cite{gou2013dynamic,yang2013dynamic,ren2014use,liang2009dynamic}. The thermo-chemical state at the $n+1$ time-step is only a function of  $Y_{k,n}^{Ret}$ and $\dot{\omega}_{Y_{k,n}^{Ret}}$, as represented here:

  \begin{equation}
     Y_{k,n+1}^{Ret} = Y_{k,n}^{Ret} + \dot{\omega}_{Y_{k,n}^{Ret}} \, dt + \tau_r (Y_{k,n}),
     \label{eq:intt}
 \end{equation}
 
 where $\dot{\omega}_{Y_{k,n}^{Ret}}$ are the chemical source terms for the retained species, $dt$ is the time-step interval and $\tau_r (Y_{k,n})$ represents the transport operator that includes advection and diffusion. As any of the species from the reference mechanism can become chemically active at different stages of the combustion process, all species are transported (i.e., $\tau_r$ is applied to all species $Y_{k}$). This is ensured by performing the chemistry reduction starting from the detailed mechanism and taking into account the chemistry of all species. Hence the reduced schemes generated locally would always be representative of the detailed mechanism.
 
  \subsection{TRAC method}

 The definition of the thermo-chemical database to store the locally reduced reaction mechanisms in TRAC is given by a discrete space in mixture fraction $Y_{\xi}$ and a chemical progress variable $Y_{C}$, respectively. While the Bilger's formula
 is used for $Y_{\xi}$, the choice of progress variable is arbitrary. The impact of the definition of $Y_{C}$ on the generation of the reduced schemes is limited. Unlike the flamelet methods, where the definition of the progress variable can have a significant influence on the description of chemistry, in TRAC the chemical source terms and transport properties are computed from local conditions using the reduced reaction mechanisms. Hence a simple definition of the progress variable such as temperature or a linear combination of mass fractions of major products is, in general, sufficient to identify evolving thermal states and their corresponding reduced reaction mechanisms. For this study, a simple definition of $Y_{C}$ based on a linear combination of major species~\cite{benajes2022analysis,GOVERT2015804,govert2018effect} is used: 
  
 \begin{equation}
     Y_{C} = \sum_{i =1}^{4} b_{i} \, \frac{Y_{k}}{W_{k}},
     \label{eq:progvar}
 \end{equation}
 
 where $Y_{k}$ and $W_{k}$ are the mass fractions and molecular weights of species $\mathrm{CO_{2}}, \mathrm{H_{2}O},\mathrm{H_{2}},\mathrm{CO}$, with the constant b=[4,2,0.5,1], respectively. 
 In this study, the aforementioned classical progress variable definition was found to obtain accurate predictions of major and minor species in unsteady multiregime conditions, but this should be revised when testing other fuels and conditions.
 The TRAC table is discretized by 31 and 51 points in $Y_{\xi}$ and $Y_{C}$, respectively. This discretization is motivated by prior experiences which showed that the uniqueness of the generated reduced mechanisms is more sensitive to variations in $Y_{C}$ than $Y_{\xi}$. The discretization is linear in $Y_{C}$ and is geometric in $Y_{\xi}$ centered about $Y_{\xi, st}$, where $Y_{\xi, st}$ is the stoichiometric mixture fraction. In the TRAC table, $Y_{\xi}$ ranges from $Y_{\xi, 1}$ to $Y_{\xi, 2}$, where  $Y_{\xi, 1}$ to $Y_{\xi, 2}$, are the minimum and the maximum $Y_{\xi}$, encountered in the problem, respectively. Additionally $Y_{C}$ ranges between 0 and the maximum of $Y_{C}(Y_{\xi})$ across all $Y_{\xi}$. All test cases involve combustion of methane-air mixtures at atmospheric pressure and the GRI 3.0 \cite{gri} containing $N_{sp} \, =  53$ species and $N_{r} \,= 325$ reactions is used as the reference mechanism for the detailed chemistry. The resulting size of the TRAC table is around 4 Mb, which is in general, a few orders of magnitude smaller than classical flamelet or ISAT tables. The memory needed for the TRAC table scales linearly with the number of points discretizing the controlling variables and the number of reactions in the detailed mechanism.

\section{Results and discussion}\label{Sec:3}\addvspace{10pt}

This section presents three fundamental problems to test the new TRAC proposal in unsteady and multiregime conditions. These cases provide critical insights into the functionality of the DAC methods in representative conditions, but without the addition of complex effects like turbulence or turbulence-chemistry interactions. The selected cases include the problem of auto-ignition of a homogeneous reactor, transient counterflow diffusion flames and a premixed flame propagating into a stratified mixture hereby referred as triple flame~\cite{knudsen2012capabilities}. The latter case does not only feature unsteady effects, but also multiregime phenomena, and can serve to test the method in more complex conditions. The analysis includes comparisons between TRAC with CODAC and detailed chemistry.

\begin{subsection}{Autoignition of a homogeneous reactor}\addvspace{10pt}
  The homogeneous reactor represents an unsteady problem dominated by chemistry and hence is first used to test the TRAC method in the absence of transport phenomena such as convection and diffusion. The reactor is assumed to evolve through partially equilibrium states at constant pressure before proceeding to a steady-state solution. The initial thermo-chemical state is defined by a stoichiometric methane/air mixture at $1500 \, K$ and atmospheric pressure. Reactions in the mixture are allowed to proceed and the temporal evolution of the thermo-chemical states is recorded and analyzed. The temperature, and species mass fractions of $\mathrm{CO}$, $\mathrm{CO_{2}}$ and $\mathrm{H_{2}O}$ for TRAC are compared with those of CODAC and detailed chemistry in Fig.~(\ref{fig:hrrT}). Truncation of certain radicals with their underlying reaction pathways causes a slight shift in the time evolution of the profiles of both DAC methods when compared to the detailed chemistry. The error is magnified by the use of RFA, however, both the magnitude and shape of the profiles correlate well with the reference solution. The error can be controlled by appropriate definition of the thresholds and an error below 2\% in ignition delay is achieved with a PFA threshold of ${.pdfilon^*_{PFA}}=0.85$.
  
The influence of RFA in terms of the reduction of reactions and associated error can be seen in Fig.~(\ref{fig:hrrpfa2}), where a contour plot representing different levels of RFA thresholds ($.pdfilon_{RFA}$) is shown.
  
\begin{figure}[H]
  \centering
  \centering
  \includegraphics[width=0.7\linewidth]{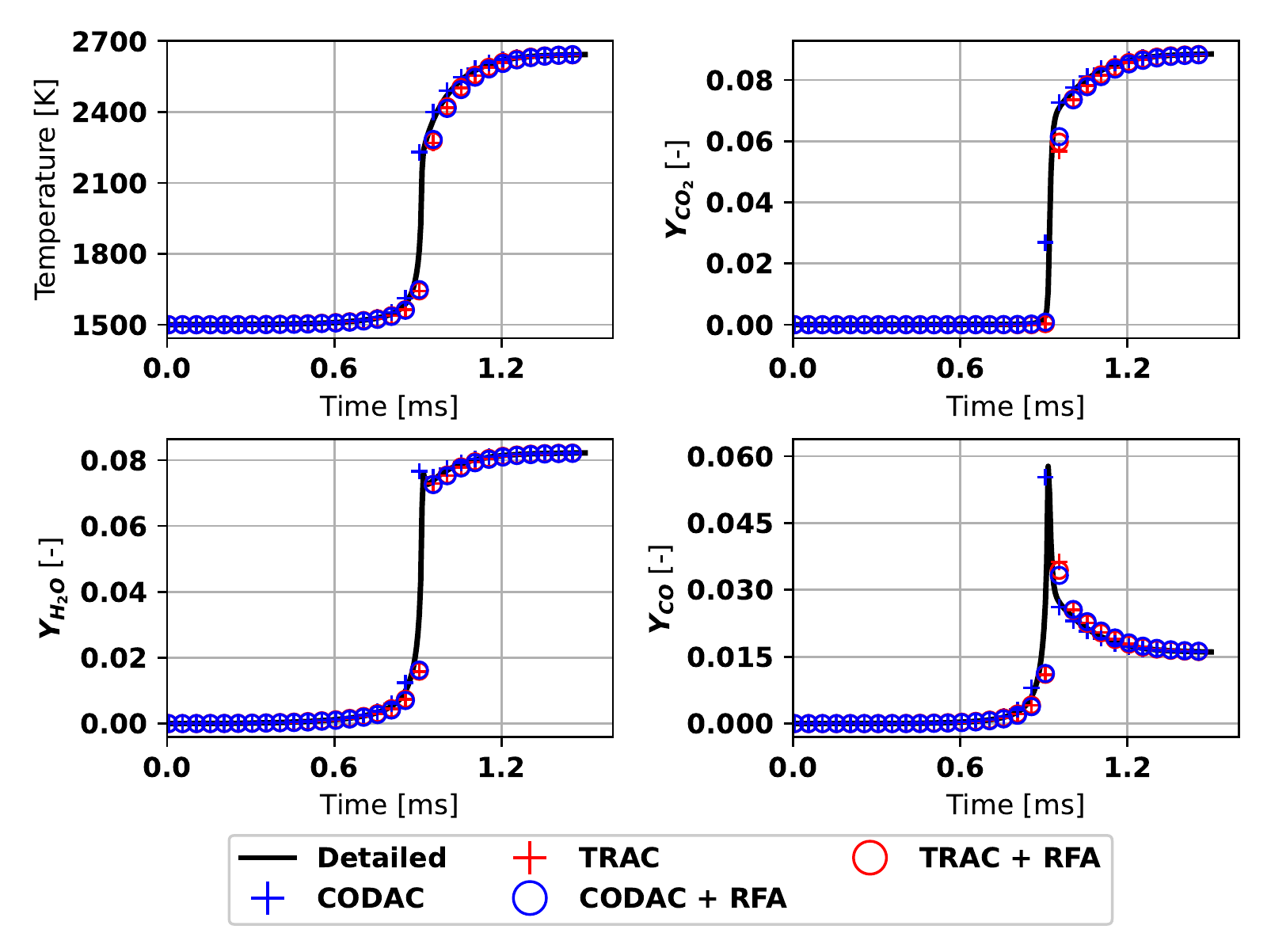}  \vspace{8 pt}
  \caption{Temporal evolution of temperature (top left), mass fraction - $\mathrm{CO_{2}}$ (top right), mass fraction - ${\mathrm{H_{2}O}}$ (bottom right) and mass fraction - $\mathrm{CO}$ (bottom left) for obtained from solutions of the TRAC and the CODAC compared to detailed chemistry.}
  \label{fig:hrrT}
\end{figure}

\begin {figure}[H]
\begin{tikzpicture}
\node[inner sep=0pt] (nreac) at (-4,0)
    {\includegraphics[width=.45\textwidth]{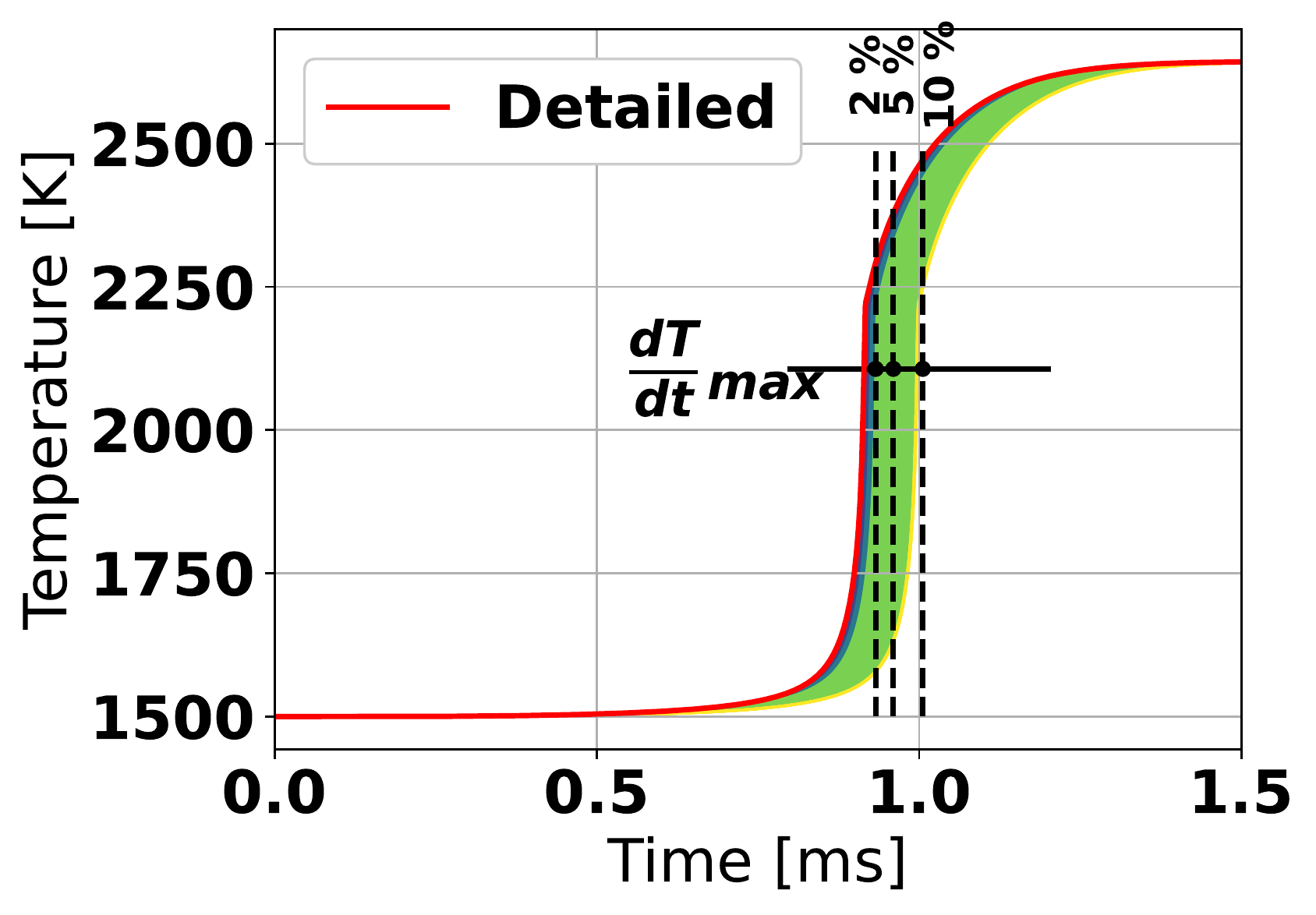}};
\node[inner sep=0pt] (nreac) at (4.0,0)
    {\includegraphics[width=.45\textwidth]{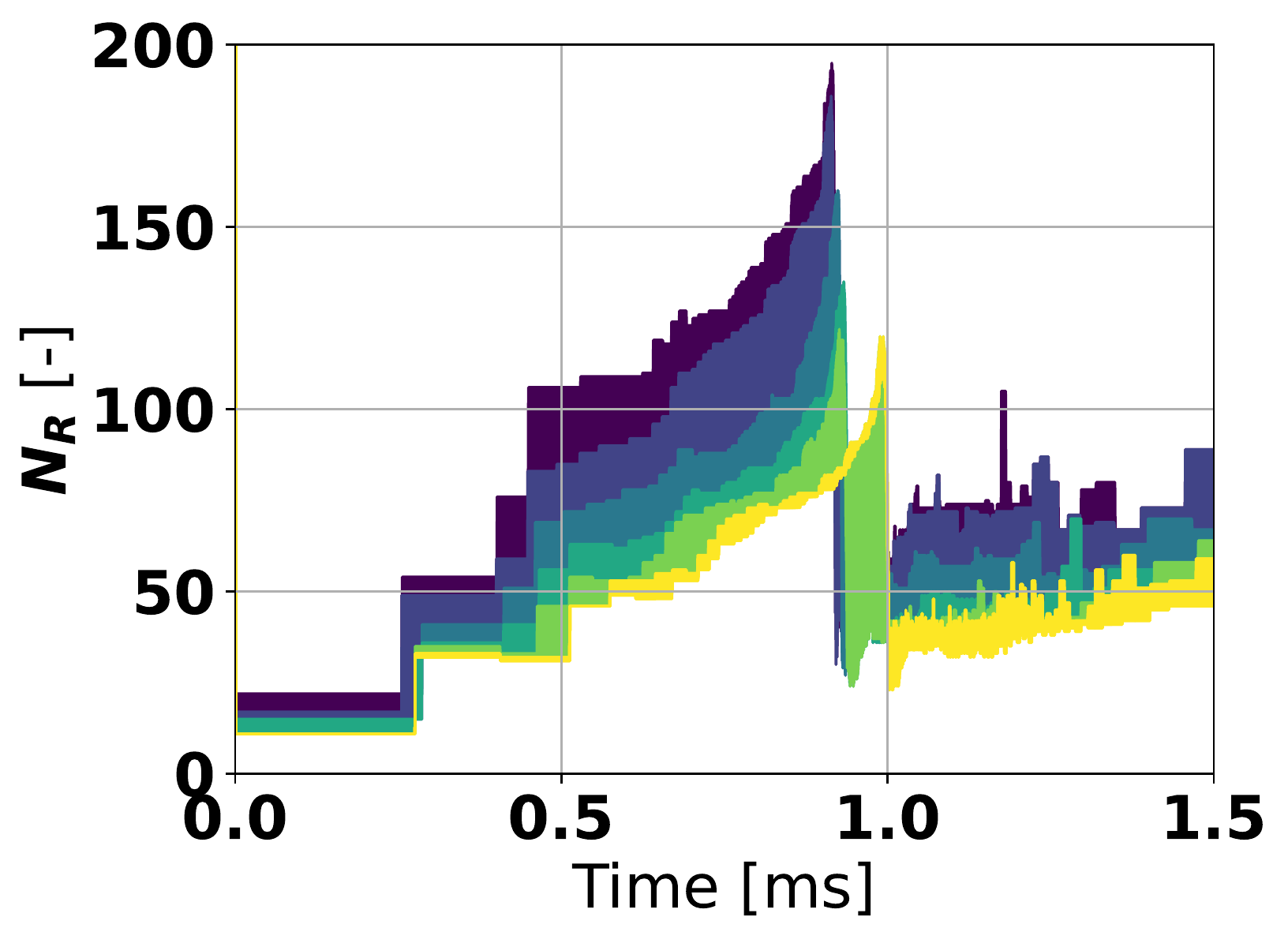}};
\node[inner sep=0pt] (nreac) at (0.0,-3.2)
    {\includegraphics[width=.65\textwidth]{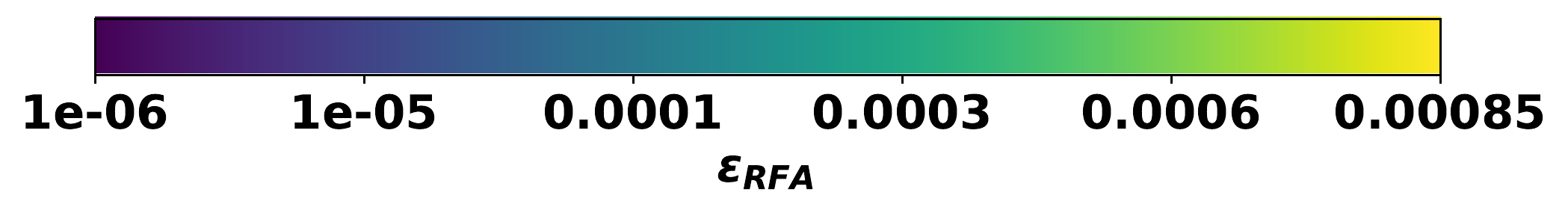}};
\end{tikzpicture}
  \vspace{8 pt}
\caption{Time evolution of temperature (left) and the number of reactions $N_{r}$ (right) for TRAC methods at various RFA thresholds with the error computed at ${(dT/dt})_{max}$.}
\label{fig:hrrpfa2}
\end {figure}

It is seen that the degree of chemistry reduction with the TRAC method correlates well with the reactivity of the mixture. The reduction is higher in the pre-and post-ignition states, where a lower number of reactions are required to describe the chemical evolution. On the other hand, the reduction algorithm identifies reaction paths with higher chemical activity in the most reactive phases of the auto-ignition process. The left-hand side plot of Fig.~(\ref{fig:hrrpfa2}) quantifies the error in auto-ignition delay due to the RFA reduction. There is a sharp increase in the auto-ignition delay error, from $2$ $\%$ to $10$ $\%$ at the RFA threshold corresponding to a reduced scheme with $N_{r} \approx 110$. A similar trend was observed during the \emph{a priori} static reduction example shown in  section 2, with an increase in error at a reduced scheme with $N_{r} \approx 200$. In section 2, a single reduced reaction mechanism was used for all states of the auto-ignition problem. However, for the same error, dynamic reduction methods with localized reduced mechanisms achieved an additional reduction of 50 $\%$. This demonstrates the advantages of using locally-reduced mechanisms to compute the chemical source terms.

\end{subsection}

\begin{subsection}{Counter flow diffusion}

The homogeneous auto-ignition case discussed in the previous section is purely dominated by chemistry, wherein the reactions proceed to chemical equilibrium at a given timescale. In such cases, the variation in thermo-chemical conditions is governed only by chemistry, however, in cases involving transport phenomena, the chemical source term is often balanced with diffusive fluxes. This interaction is usually described by the scalar dissipation rate of mixture fraction, which influences both the transport and chemistry of the problem. 
To study the applicability of TRAC in problems where the chemical source is limited by diffusion, a counter flow diffusion flame configuration is selected, see Fig.~(\ref{fig:oppnew2}). In a counter flow problem, fuel and oxidizer inlets are separated by a distance, and after a source of ignition, a diffusion flame is formed around the stoichiometric point between fuel and oxidizer. The rates at which reactants are supplied to the reacting zone has a strong influence on the burning rates, this dependency poses a challenge to the reduced reaction mechanisms to reproduce such behaviour. 

To ensure ease of computations, strain is varied by changing the mass flow rates of the inlet, while ensuring equal momentum between the inlets to maintain the stagnation plane in the middle of the domain. The 2D computational domain is defined with a gap distance $g= 12.5$ mm, stream diameters $d = 2.0$ g and domain length of $L = 2.8 \, g$. The mesh is non-uniform in the longitudinal direction and centered around the center line of the inlets. This resolution ensures the both chemical and flow scales are well captured in the areas of interest. 

 \begin{figure}[H]
  \centering
  \includegraphics[scale=0.5]{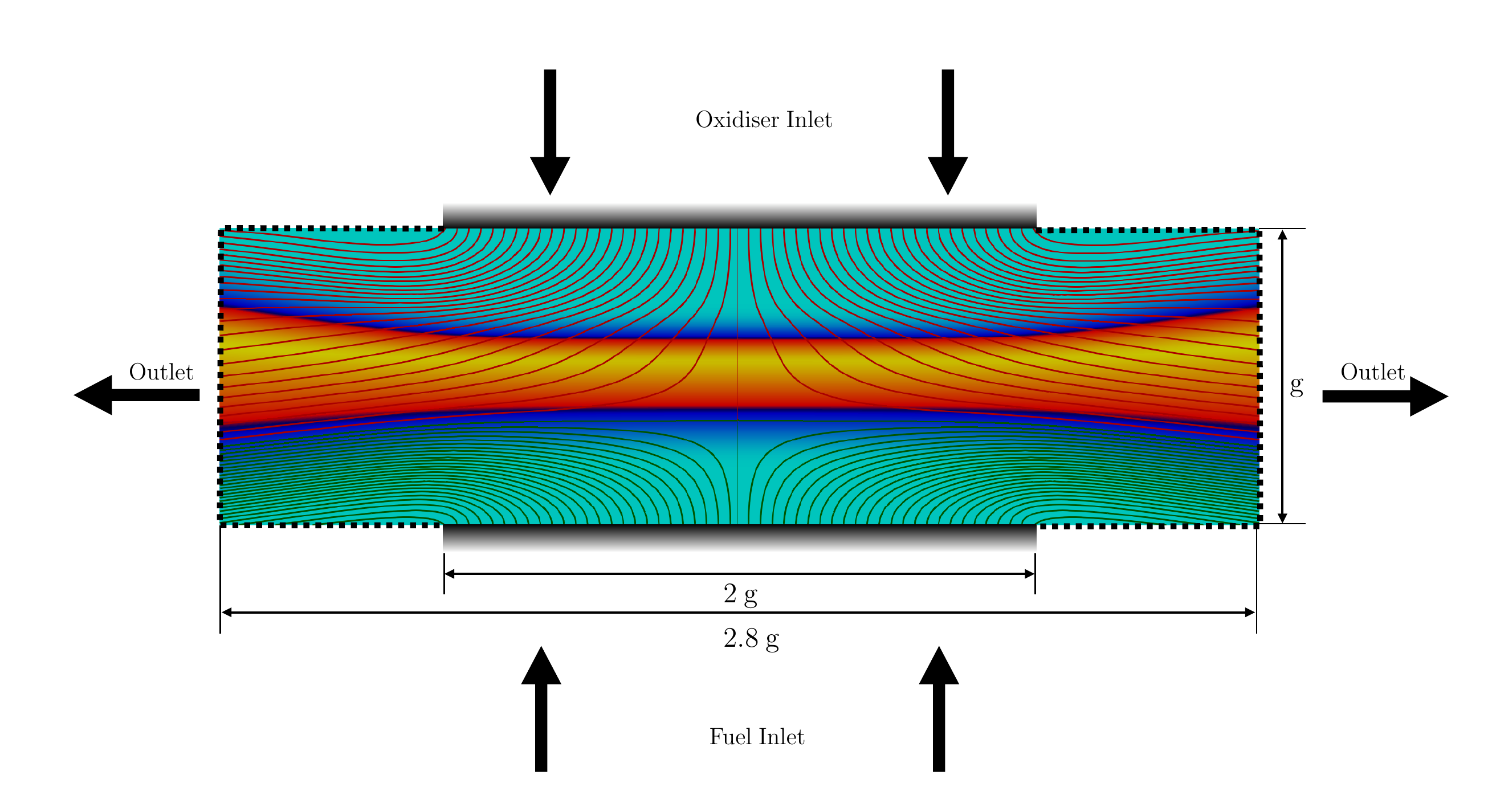}
  \caption{Opposite jet counter flow diffusion schematic. Arrows indicate the inlets and the outlet, respectively and the dotted line indicates Neumann boundary conditions for velocity and scalars.}
  \label{fig:oppnew2}
\end{figure}

In the subsequent sections, two transient problems with this counter flow configuration are presented: 1) extinction under strain and 2) auto-ignition under strain. Details of the different cases are given in the next subsections.

\begin{subsubsection}{Extinction under strain}

Extinction by strain is a common phenomenon in many applications, ranging from jet flames like the series of Sandia D-F flames to spray flames~\cite{benajes2022analysis,garmory2011capturing}. This phenomenon is characterized by a rapid evolution of a flame from stable operation to extinction after an increase in the strain level.  Extinction is produced when the mixing rates are higher than the time scales of reactions, so combustion can no longer be sustained and the flame extinguishes. This exact process is reproduced here using numerical simulations in a counter flow flame configuration. From a stable flame at strain $s= 30$ 1/s, a strain rate increase of $\Delta s / s = 3.5 $ is imposed by changing the mass flow rates accordingly. Results are recorded on the  center line in time and are presented below. 

As the strain increases, more reactants are forced away from the stagnation plane at rates faster than the burning rate, which leads to the reaction front having a lower volume and a subsequent increase in the local heat release rate. The resulting effect is, however, an overall decrease in the total heat release and temperature, which continues to a state where the reactions can no longer be sustained and the reacting front extinguishes. A graphical representation of the extinction process for the temperature and heat release rate at different time instants is shown in Fig.~(\ref{fig:ext3333}).

The scalar dissipation rate ($\chi$), which gives a measure of the molecular mixing and is directly proportional to the strain is used to further study the extinction process. The maximum temperature and the scalar dissipation rate along the center line are shown in Fig.~(\ref{fig:ext11}(a)). There is a continuous, almost linear increase in $\chi$, corresponding to a similar decrease in the maximum temperature until around $\chi$ = 60 1/s, where there is a steep drop in temperature indicating flame extinction. Similar trends are observed in the number of reactions $N_{r}$, where there is a sudden drop after the onset of extinction, see Fig.~(\ref{fig:ext11}(b)). This indicates that both TRAC and CODAC take into account the changing dynamics of the flame and adapt the chemistry accordingly.
 
\begin{figure}[H]
 \captionsetup{justification=centering}
 \begin{tabular}{cc}
   \includegraphics[width=0.45\textwidth]{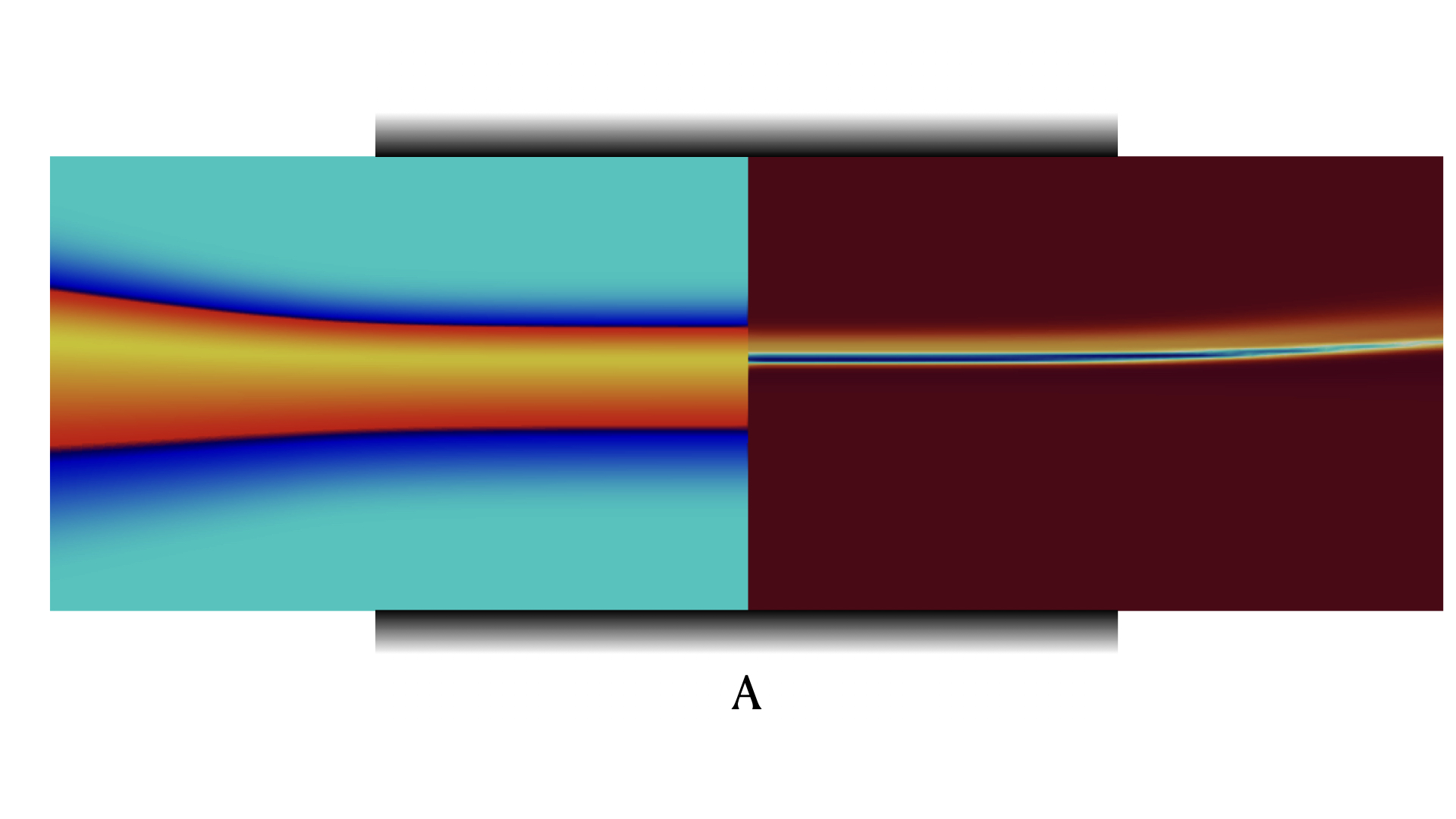}
   \includegraphics[width=0.45\textwidth]{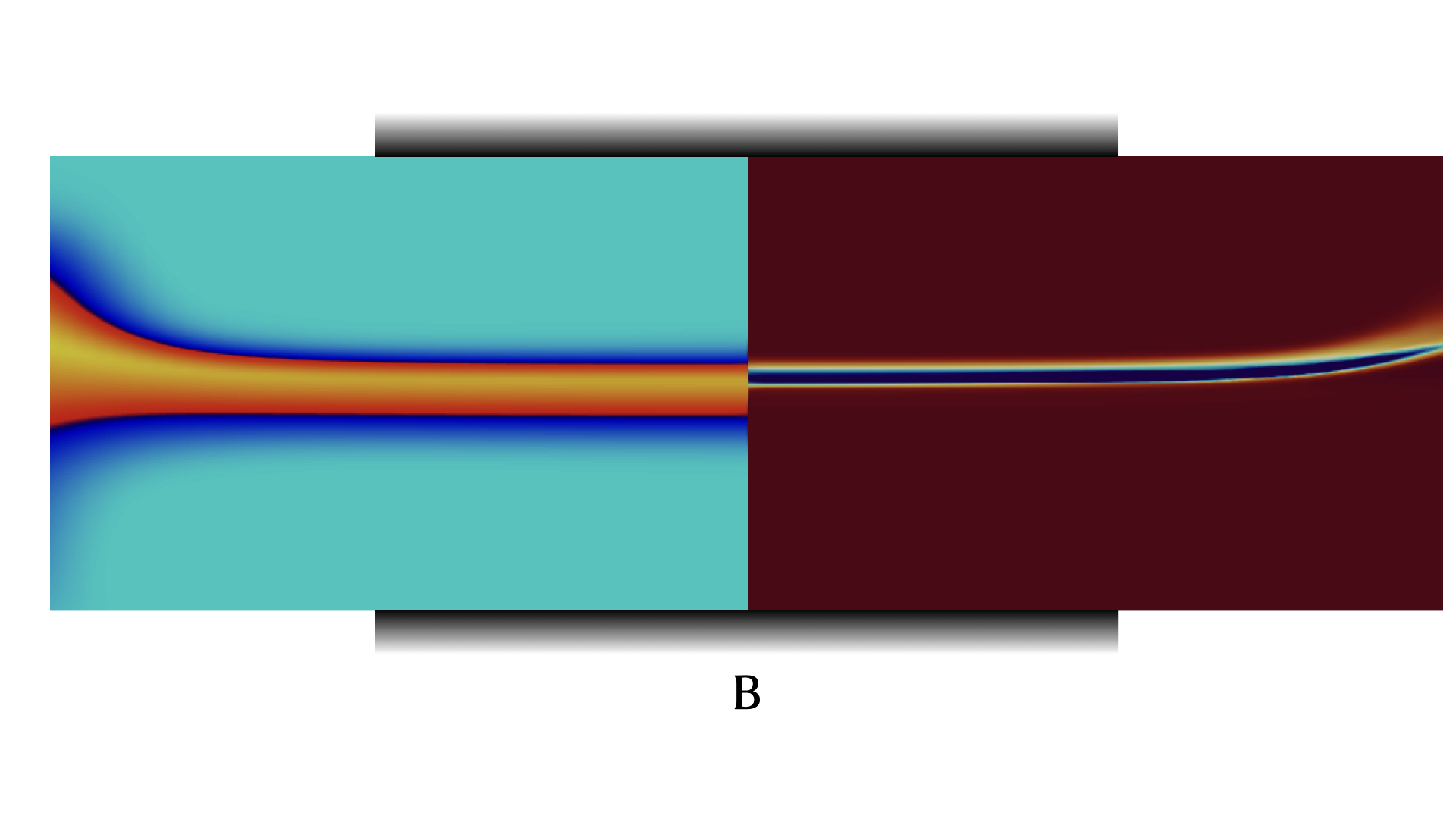} \\
   \includegraphics[width=0.45\textwidth]{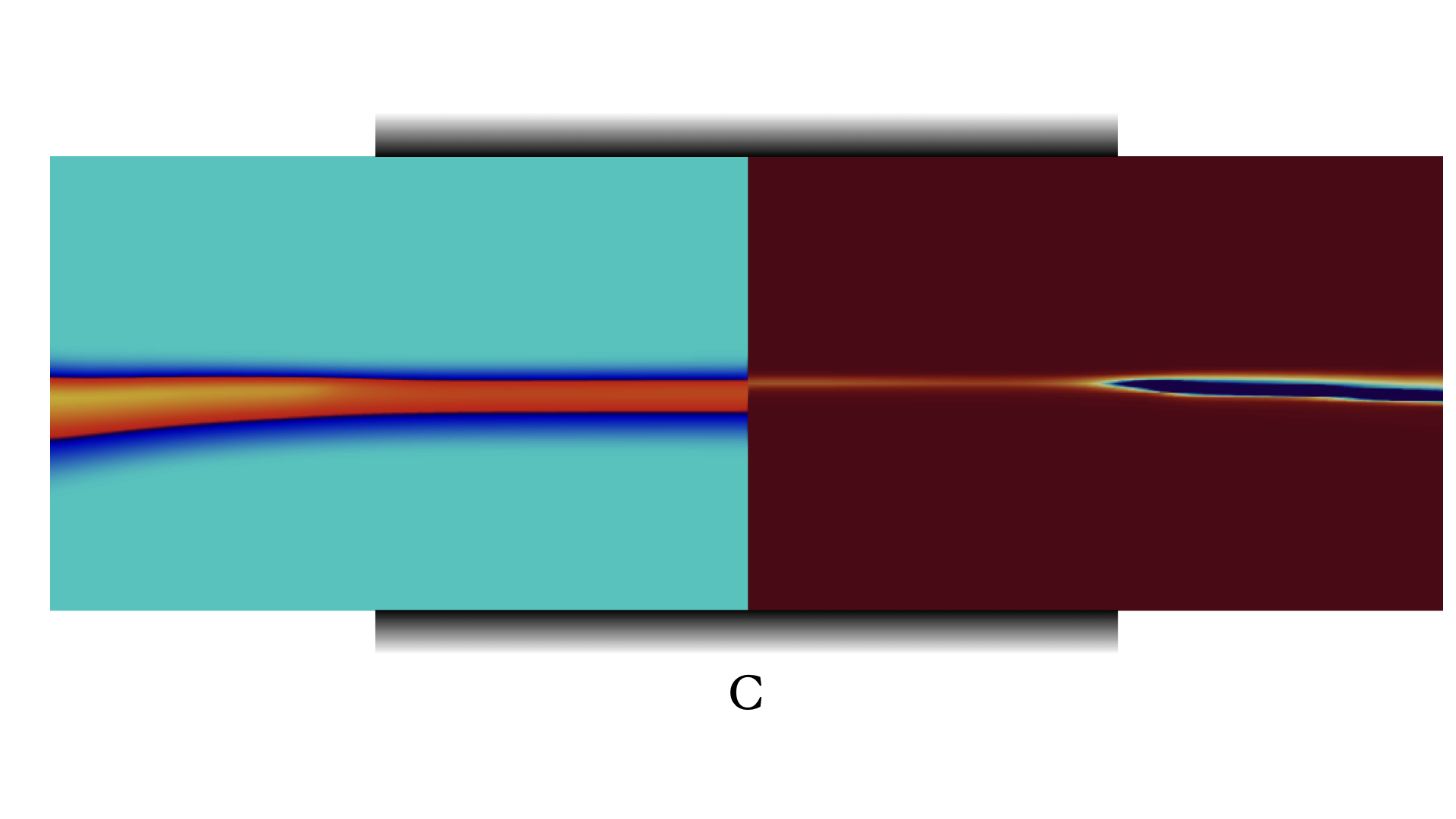}
   \includegraphics[width=0.45\textwidth]{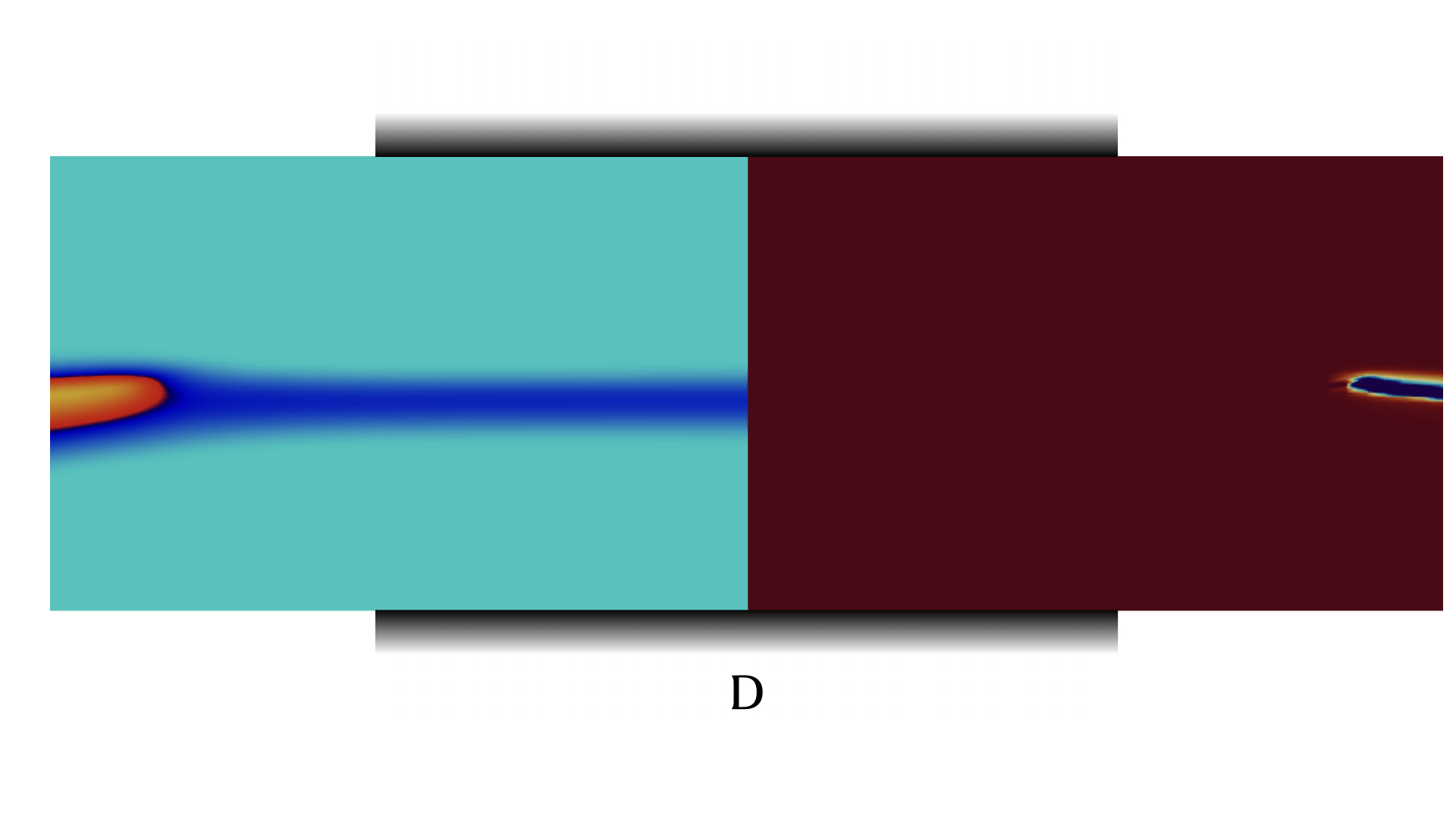} \\
   \includegraphics[width=0.45\textwidth]{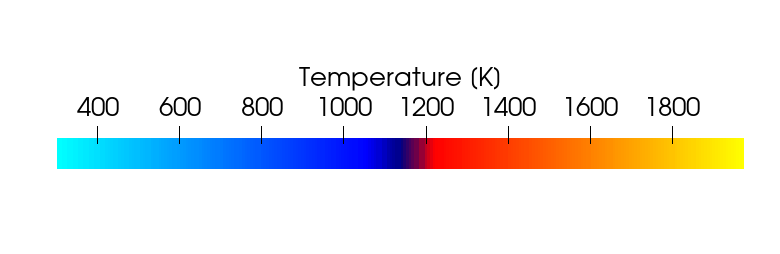}
   \includegraphics[width=0.45\textwidth]{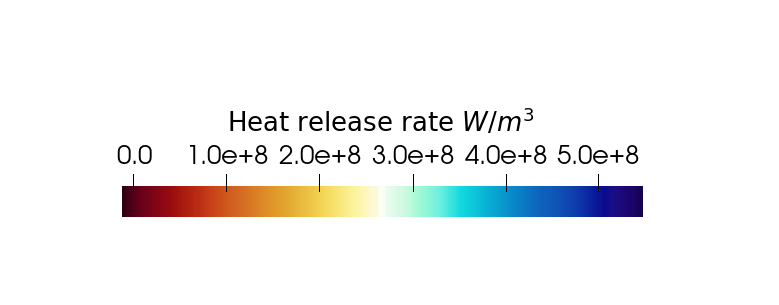}
 \end{tabular}
 \caption{Temporal evolution of extinction under strain, temperature (left half),  Heat Release Rate ($HRR$) (right half). The states are named A to D and are obtained at a physical time of 0.0, 2.2, 4.4, and 6.6 milliseconds, respectively. When expressed as normalized time $t = E (1/a)$ in relation to the extinction process, time instances correspond to the following values of E = [0.0, 0.3, 0.6, 0.9] respectively.}
 \label{fig:ext3333}
\end{figure}

\begin{figure}
\centering
\begin{subfigure}{.5\textwidth}
  \centering
  \includegraphics[width=1.0\linewidth]{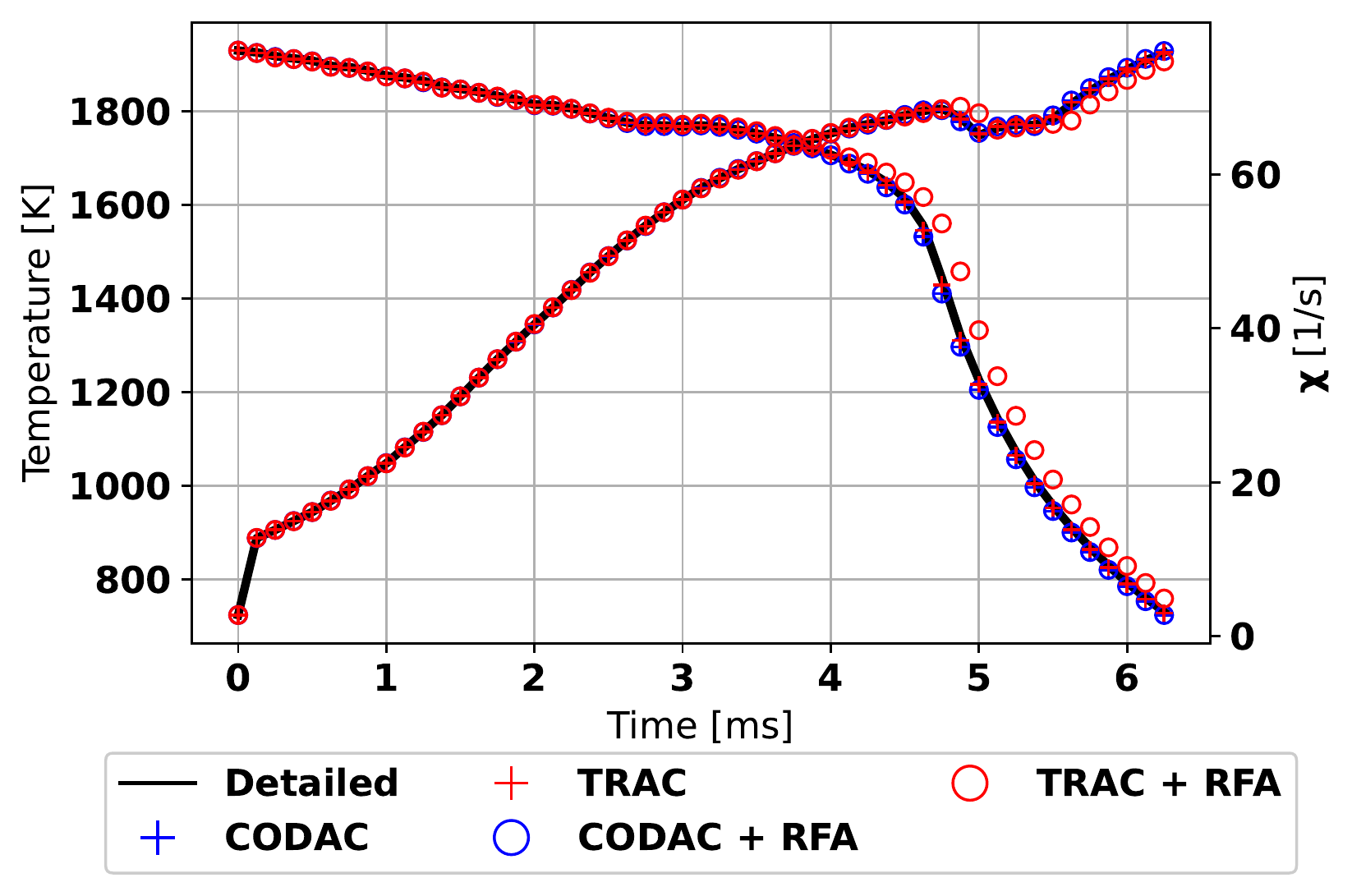}

\end{subfigure}%
\begin{subfigure}{.5\textwidth}
  \centering
  \includegraphics[width=1.0\linewidth]{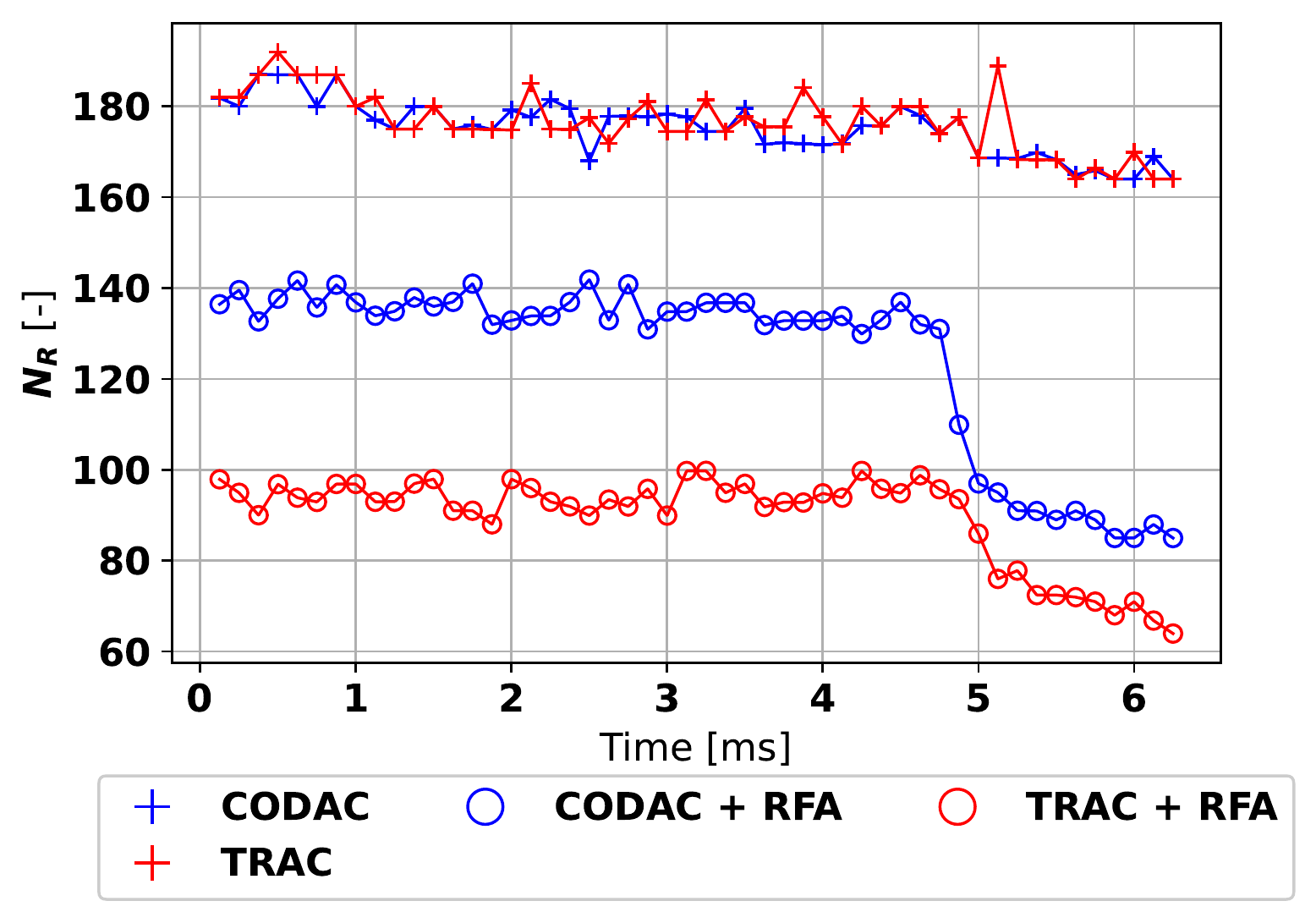}
\end{subfigure}
\caption{Maximum temperature and scalar dissipate rate $\chi$ (left) and maximum number of reactions $N_{r}$ (right) with time, obtained along the center line during the extinction process.}
\label{fig:ext11}
\end{figure}

Reduction in the chemical source terms further simplifies the chemistry of the problem, as more reactions fail to reach partial equilibrium and lead to the extinction of the flame. Solutions in mixture fraction ($Y_{\xi}$ ) space are shown in Fig.~(\ref{fig:ext22}) for TRAC compared to the reference solution. TRAC was able to predict global quantities with the locally reduced reaction mechanisms. The evolution of $\mathrm{CO}$ mass fraction is shown in Fig.~(\ref{fig:ext22}(b)). The structure of the counter flow diffusion flames is a function of both $Y_{\xi}$ and $\chi$. There are species like $\mathrm{CO}$ that show high sensitivity to both parameters~\cite{chan1998structure}, though there are other species with lower dependency like $\mathrm{H_{2}O}$. It has been shown that tabulated methods, such as ISAT with skeletal mechanisms, can fail to predict $\mathrm{CO}$ concentrations accurately during extinction~\cite{tang2000probability}. However, as TRAC and CODAC take into account the local changes in composition to identify the appropriate reduced reaction mechanisms, accurate predictions of $\mathrm{CO}$ are expected despite $\mathrm{CO}$ not being part of the key species definition in the PFA method. This can be distinguished in Fig.~(\ref{fig:ext22}(b), where a good correlation with the strain is found for TRAC using PFA. A minor exception to this trend is the slight deviations in temperature and mass fraction of $\mathrm{CO}$ from the detailed chemistry solution using the TRAC + RFA method. The phenomenon of extinction is characterized by a continuous decrease in both the consumption and production rates of species, which leads to the over-reduction of chemistry. As seen in Fig.~(\ref{fig:ext22}(b) the TRAC + RFA method uses reduced schemes which are 50 $\%$ smaller than the CODAC + RFA method, this is a consequence of the lack of sensitivity of the progress variable to extinction process and the aggressive reduction of RFA. However, the accuracy of the method can be improved by the use of additional controlling variables or by tuning the RFA threshold accordingly. Furthermore the RFA method can be modified to use integrated chemical sources rather than instantaneous to account for history effects as shown in \cite{rfabook}.

\begin{figure}[H]
\centering
\begin{subfigure}{.5\textwidth}
  \centering
  \includegraphics[width=1.0\linewidth]{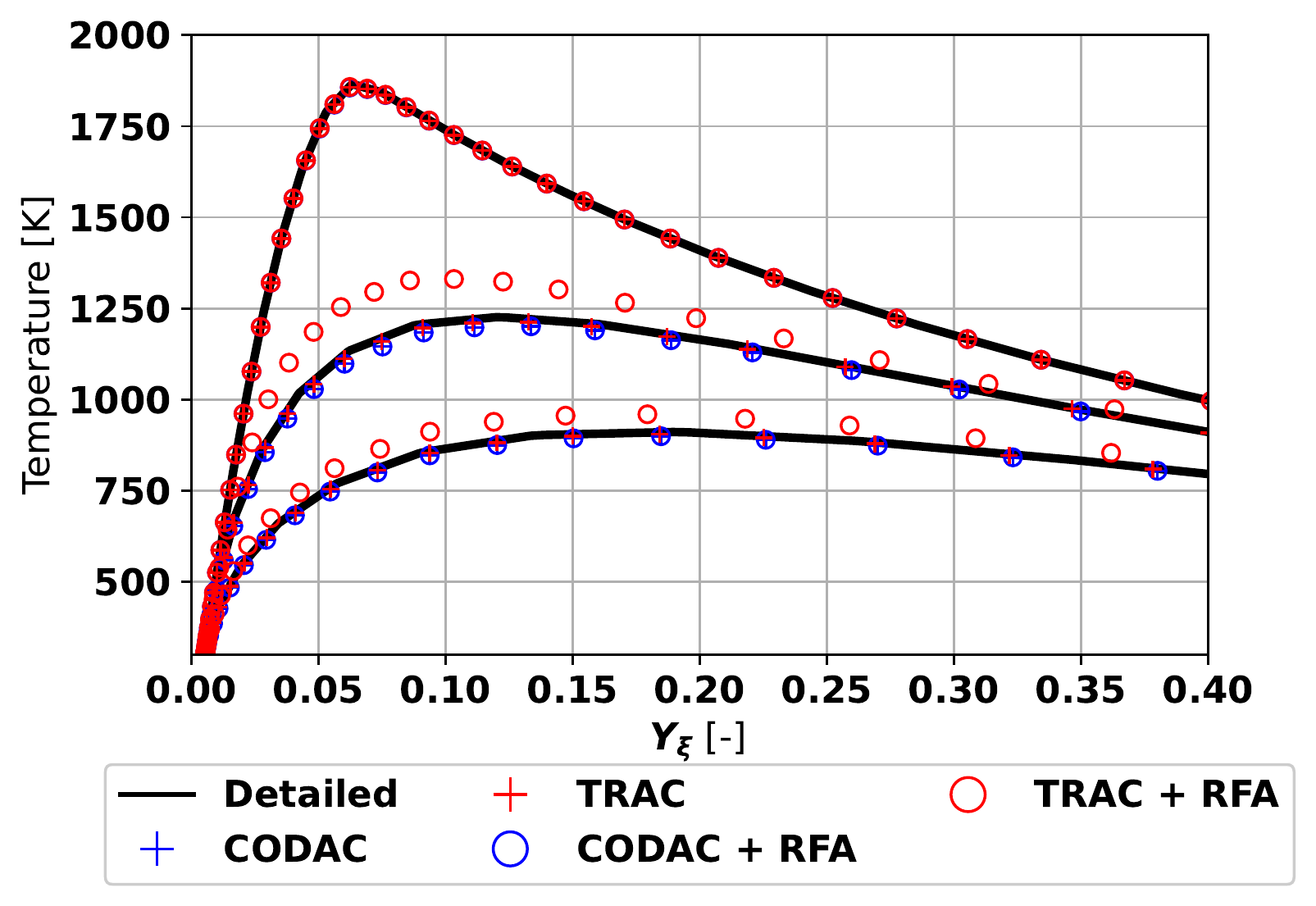}
  \label{fig:sub1}
\end{subfigure}%
\begin{subfigure}{.5\textwidth}
  \centering
  \includegraphics[width=1.0\linewidth]{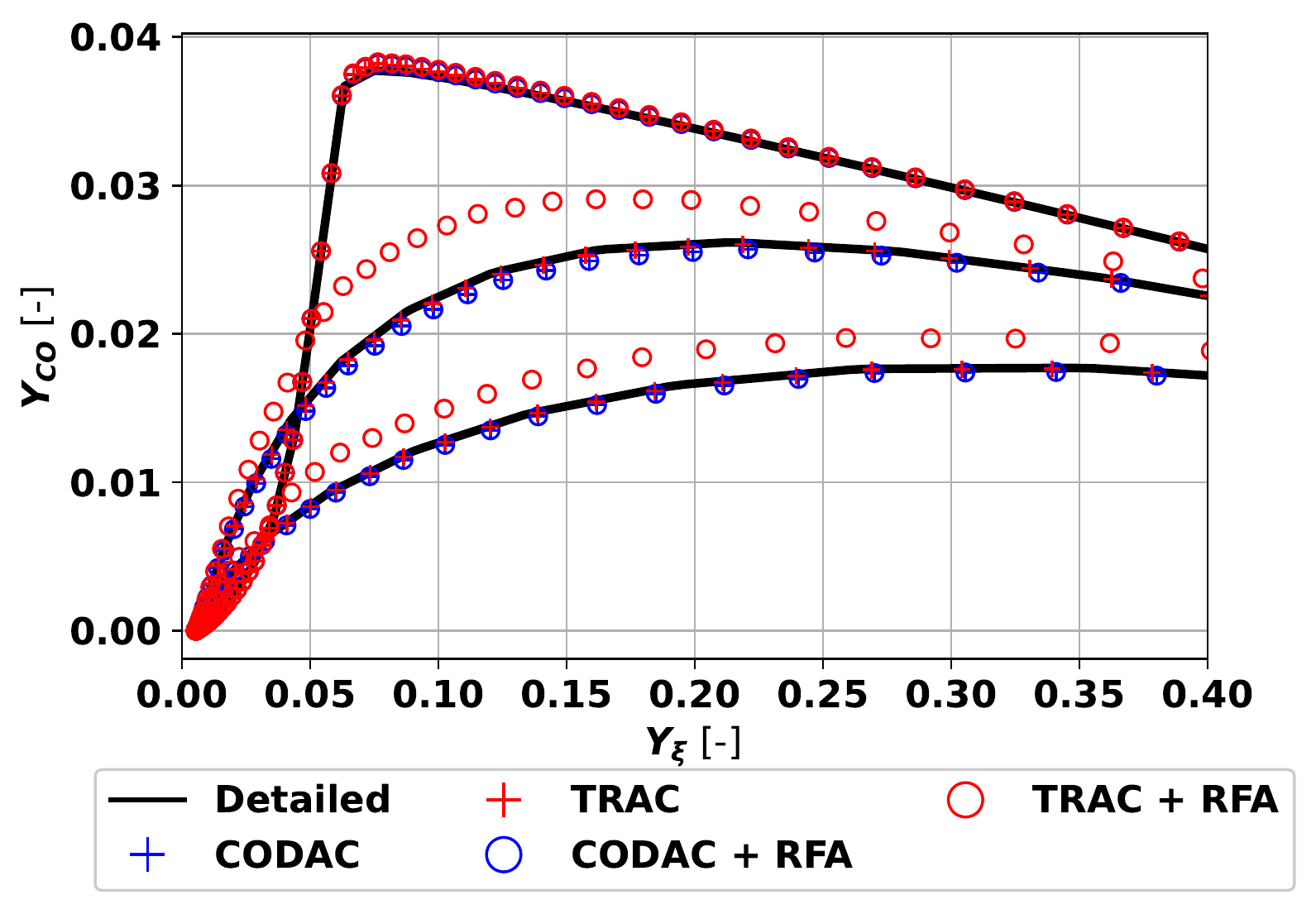}
  \label{fig:sub2}
\end{subfigure}
\caption{Extinction profiles temperature (left) and mass fraction - $\mathrm{CO}$ (right) against mixture fraction $Y_{\xi}$ obtained along the centerline at time t = $1.25, 4.375$ and $6.25$ milliseconds, respectively.}
\label{fig:ext22}
\end{figure}

\end{subsubsection}

\begin{subsubsection}{Ignition under strain}
Ignition and re-ignition are complex phenomena that occur in a wide variety of practical combustion problems. The stability of the flame is dependent on the mixing and the chemical scales leading to either stable flames in case of balanced sources or unsteady  flames (extinguishing or igniting) in cases of imbalance. In the previous section, an extinction event was investigated. It was shown how the chemical source terms vanished due to the high strain rate in the flow. In this case, the ignition process is investigated where the chemical sources are balanced by the mixing fluxes. The computational domain for these is similar to the one used in the previous section, the only difference being the initial condition, which, here was defined with a mixing solution of methane at 298 K and air at 1200 K with a global strain rate of 32 1/s. A stable flame is obtained when the diffusive fluxes balance the chemical sources, which can be distinguished by looking at the evolution of the profiles for temperature and species.

\begin{figure}
\centering
\begin{subfigure}{.5\textwidth}
  \centering
\includegraphics[width=1.0\linewidth]{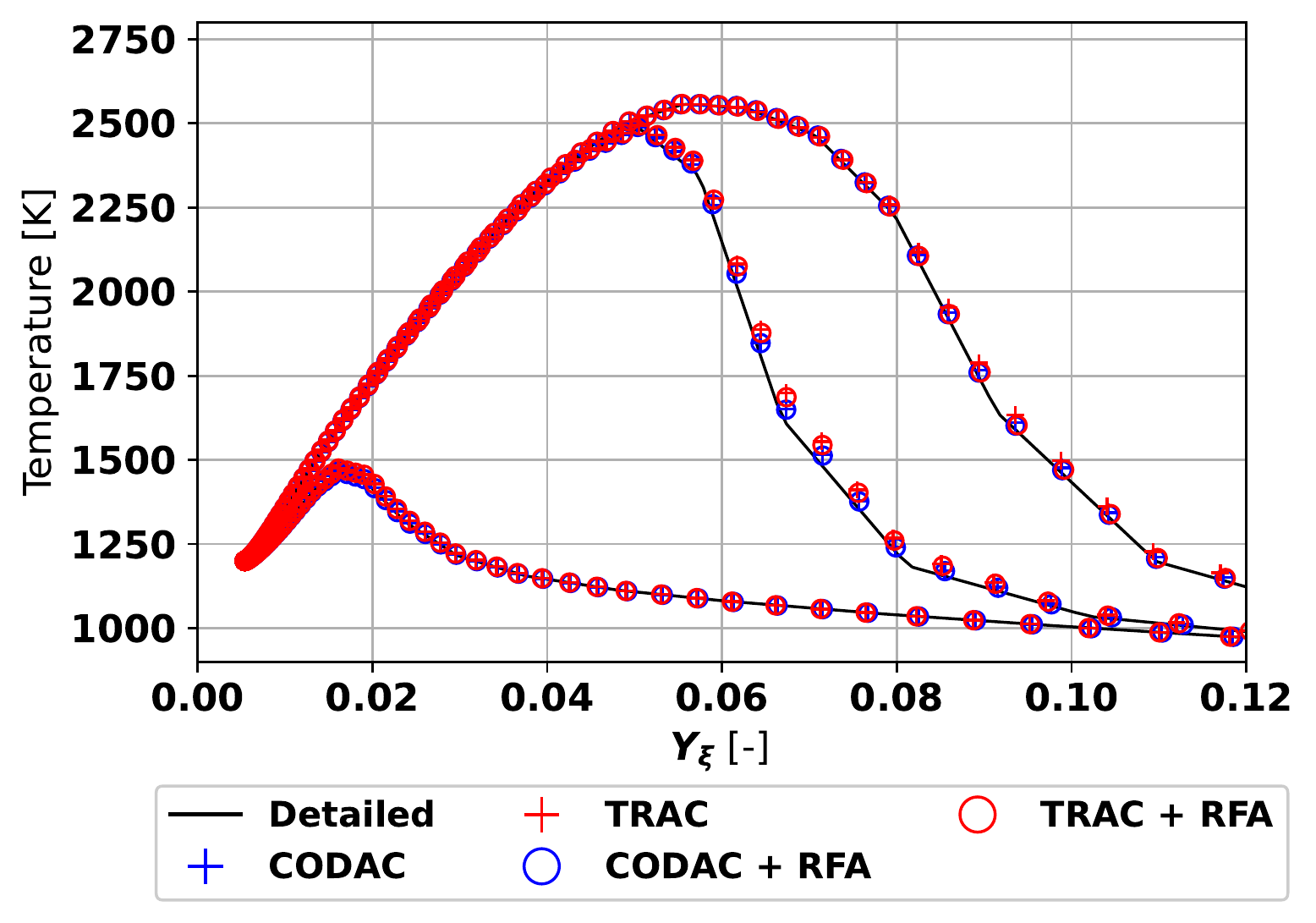}
\end{subfigure}%
\begin{subfigure}{.5\textwidth}
  \centering
  \includegraphics[width=1.0\linewidth]{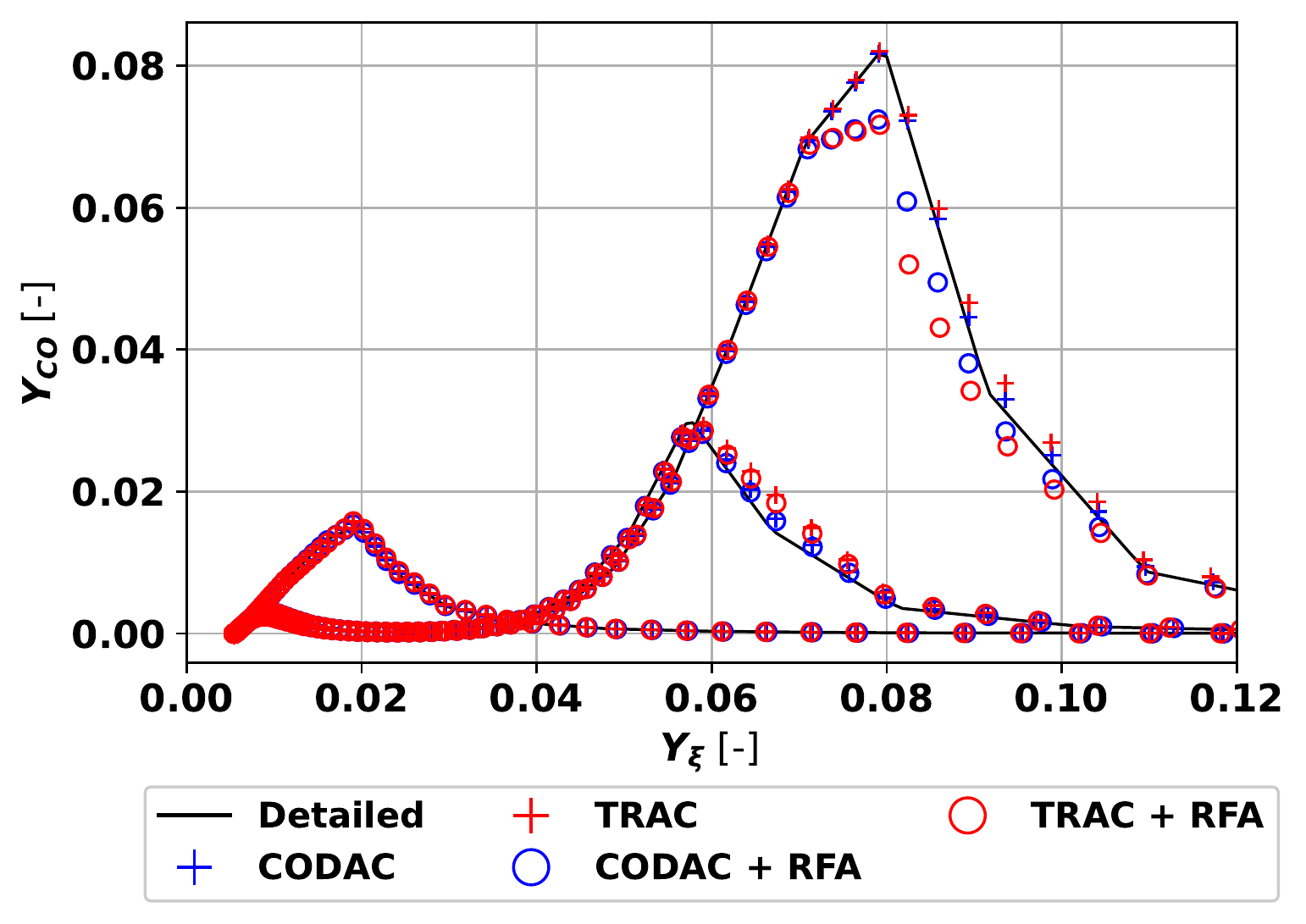}
\end{subfigure}
\caption{Ignition profiles temperature (left) and mass fraction - $\mathrm{CO}$ (right) against mixture fraction $Y_{\xi}$ obtained along the centerline at time t = $2.0, 2.55$ and $2.75$ milliseconds, respectively.}
\label{fig:ign22}
\end{figure}

\begin{figure}[H]
\centering
\begin{subfigure}{.5\textwidth}
  \centering
\includegraphics[width=1.0\linewidth]{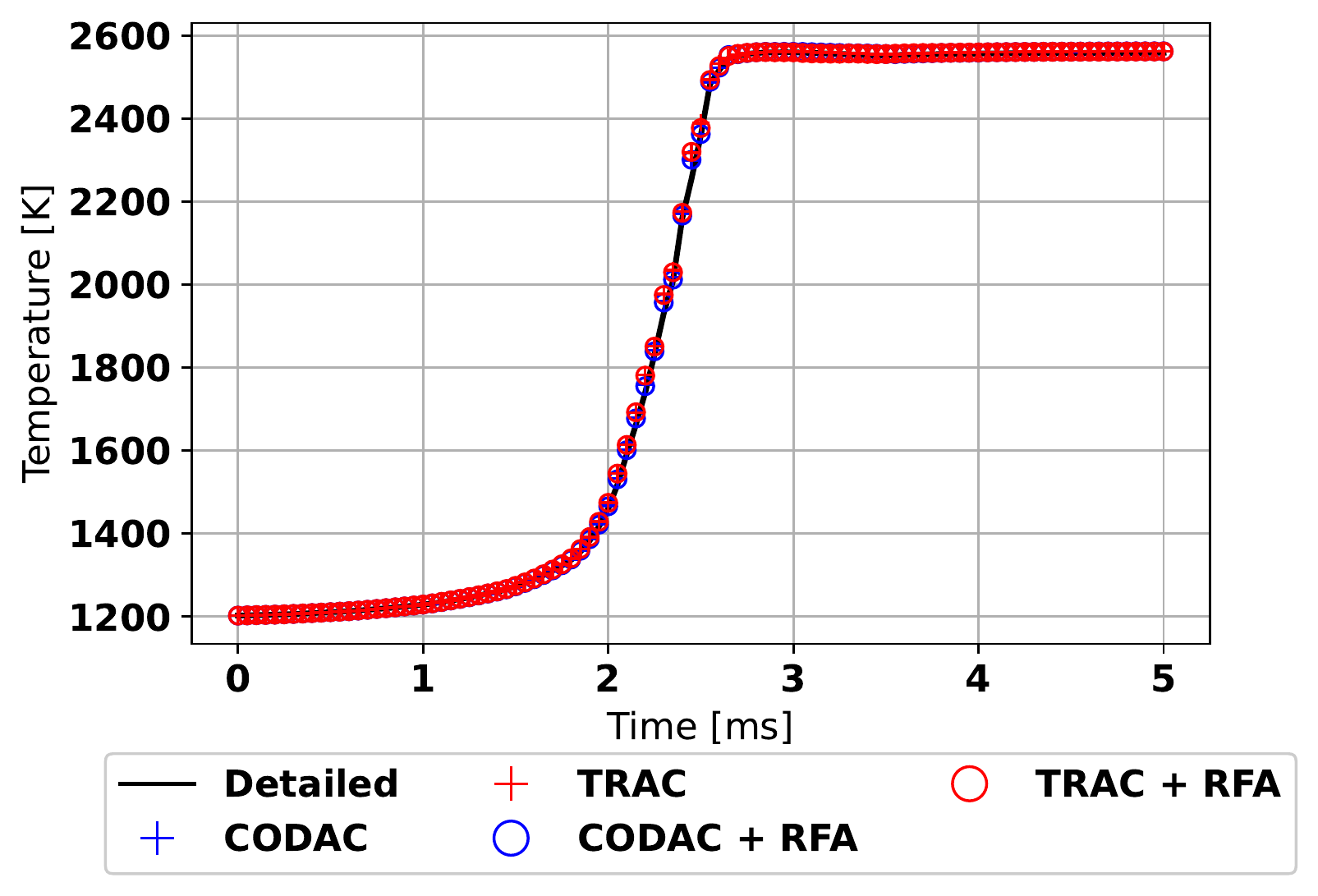}
\end{subfigure}%
\begin{subfigure}{.5\textwidth}
  \centering
  \includegraphics[width=1.0\linewidth]{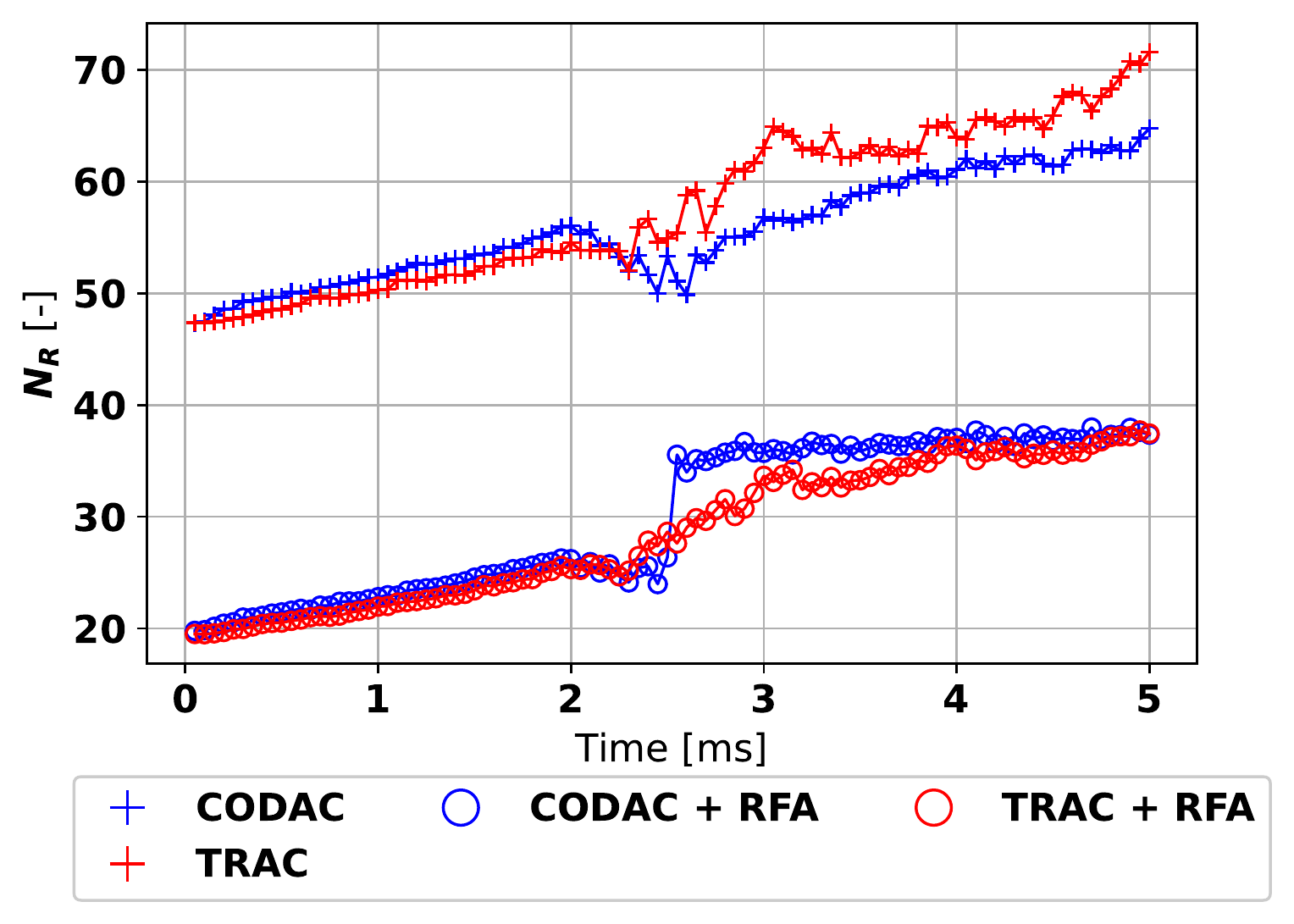}
\end{subfigure}
\caption{Maximum temperature (left) and maximum number of reactions $N_{r}$ (right) with time, obtained along the centerline during the ignition process.}
\label{fig:ign111}
\end{figure}

The temperature and $\mathrm{CO}$ mass fraction during different phases of an auto-ignition problem are shown in Fig.~(\ref{fig:ign22}) as function of mixture fraction $Y_{\xi}$. The ignition process starts in leaner mixtures and slowly proceeds to stoichiometry, as observed by the shift in the peak of the profiles of both temperature and $\mathrm{CO}$ mass fraction. Since the reduction is dynamic and based on the relative weight of the reaction pathways, the thermal state of the flame is accurately reproduced along the distinct phases. The time evolution of the maximum temperature and the number of active reactions ($N_{r}$) for this case is shown in Fig.~(\ref{fig:ign111}). As the temperature increases, the mixture becomes more reactive and requires more detailed chemical description, so the number of reactions increases. For the given strain of $32$ 1/s, the autoignition delay time is around 2.75 ms, and the radical pool is established during the first 1.5 - 2.0 ms with the advancement of chain initiation reactions. This radical pool then proceeds to chain branching reactions, which results in a rapid increase of temperature, as seen in Fig.~(\ref{fig:ign111}).

\end{subsubsection}
\end{subsection}

\begin{subsection}{Triple Flame}\addvspace{10pt}

The flame stabilization mechanism of a premixed flame front interacting with a stratified mixture is a representative condition of practical systems and is a common benchmark to validate partially premixed combustion models~\cite{illana2021extended,knudsen2012capabilities}. Three distinct phases of combustion can be observed in these flames, a premixed flame near stoichiometry, a partially premixed flame close to the leading edge, and a diffusion flame at the trailing end. This configuration is proposed to evaluate the impact of the chemical error on the thermo-chemistry of the flame, and hence, on the flame burning velocity.

\begin{figure}[H]
\centering
\begin{subfigure}{.75\textwidth}
  \centering
  \includegraphics[width=0.9\linewidth]{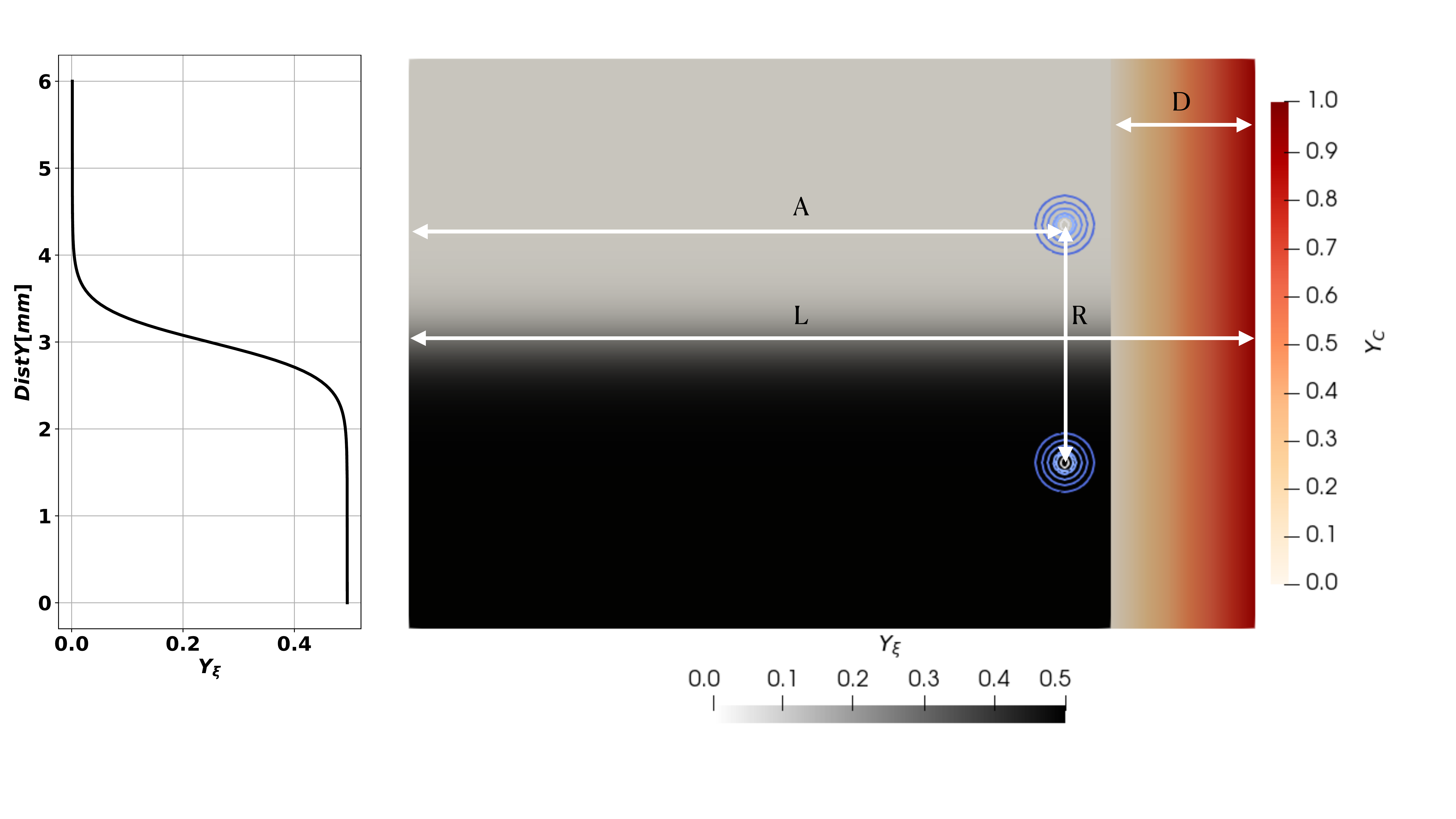}
\end{subfigure}
  \vspace{8 pt}
\caption{Triple flame schematic, adapted from \cite{knudsen2012capabilities}.}
\label{fig:tri}
\end{figure}

The triple flame configuration studied here is taken from the work of Knudsen et al. \cite{knudsen2012capabilities}. The case is set in a domain of size $[{L},2{L}/3]$ with $L=9$ mm and a two-dimensional quadrangular mesh of 300 and 150 points in the X and Y directions, respectively. The initial fields namely, the gradient of the mixture fraction and tangential velocity of the counter-rotating vortices were defined according to Knudsen et al. \cite{knudsen2012capabilities} but adapted to methane. This problem is defined by a flow mixture coming from the left boundary with a constant inlet velocity  of $u=0.32$ m/s and with a composition that prescribes the specific mixture fraction gradient in the vertical direction. The mixture fraction distribution is a hyperbolic tangent. Two counter-rotating vortices separated by $R = L/3 $ are imposed at a distance $A = 5L/6$ from the inlet plane at the start, where the composition is fixed to satisfy the target stratification. A perfectly premixed flame front is imposed at the center of the counter-rotating vortices downstream of the flow at distance $D = L/6$, which is represented in red on the sketch. After this initial field, the flame travels upstream facing the stratified mixture and forms a transient triple flame that eventually stabilizes as the vortices are convected out. The temporal evolution of the triple flame is highly dependent on the ratio of the longitudinal dimension ($\Theta_{L}$) of the domain to the flame thickness ($\delta_{L}$). This ratio determines the magnitude of the stratification of the flame. For this specific setup, the ratio ($\Theta_{L}$/$\delta_{L}$) was $13.6$, which corresponds to a high level of stratification. A schematic of the triple flame problem is shown in Fig.~(\ref{fig:tri}). The simulation proceeds to a steady state after the vortices are diffused and the incoming velocity interacts with the advancing flame front. Steady-state and transient characteristics of the flame are used as markers for evaluating the performance of the DAC methods.

\end{subsection}

\begin{figure}[H]
\centering
\begin{subfigure}{.75\textwidth}
  \centering
  \includegraphics[width=1.0\linewidth]{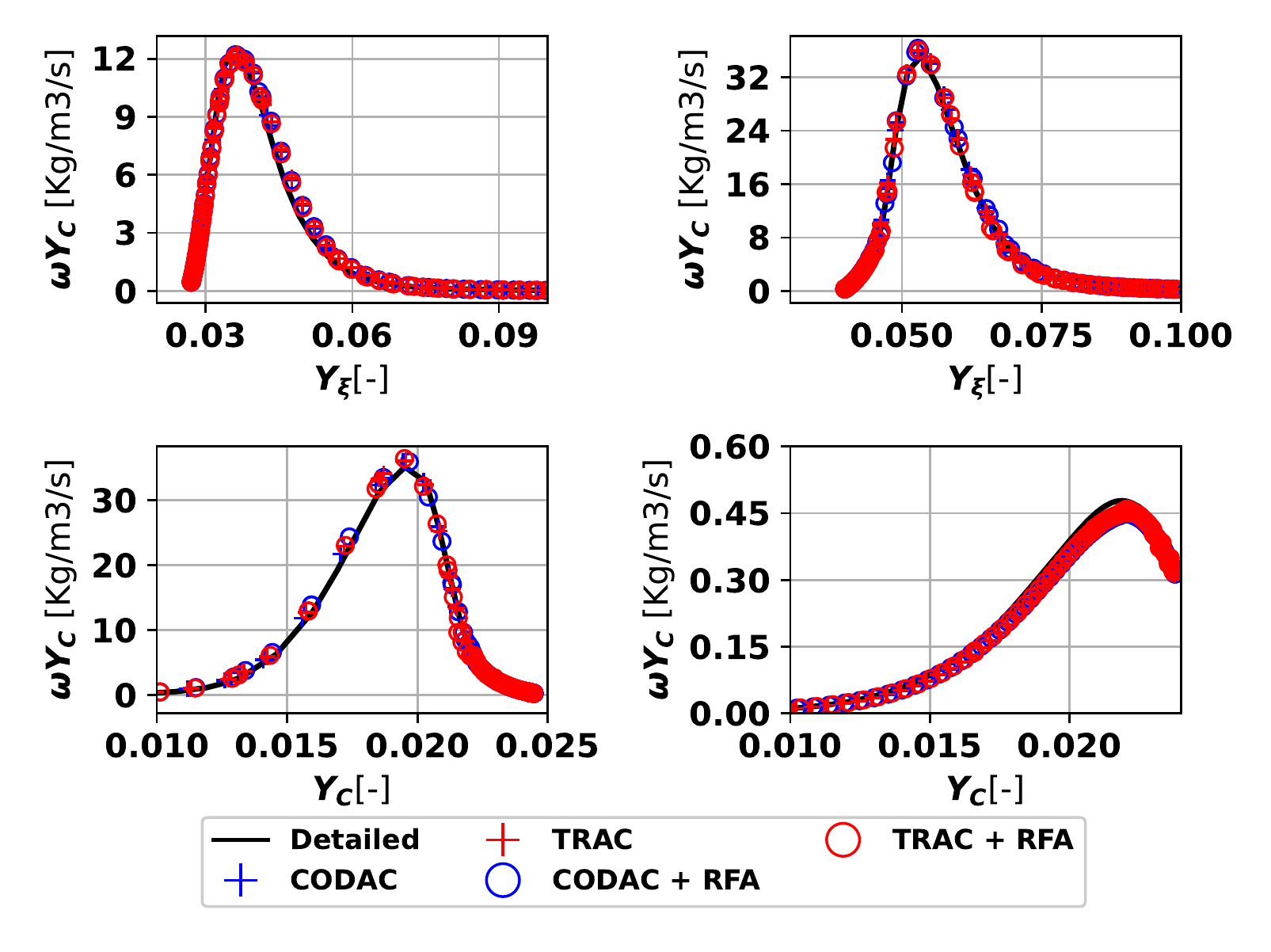}
\end{subfigure}
  \vspace{8 pt}
\caption{Source of progress variable ($\dot{\omega} Y_{c}$) along mixture Fraction ($Y_{\xi}$) at iso-line $C = 0.5$ (top left) and $C = 0.7$ (top right). Source of progress variable ($\dot{\omega} Y_{c}$) along progress variable ($Y_{c}$) at iso-line $\phi = 1$ (bottom left) and $\phi = 0.7$ (bottom right).}
\label{fig:tripleReeess}
\end{figure}

\begin{figure}[H]
 \centering
 \includegraphics[width=0.65\linewidth]{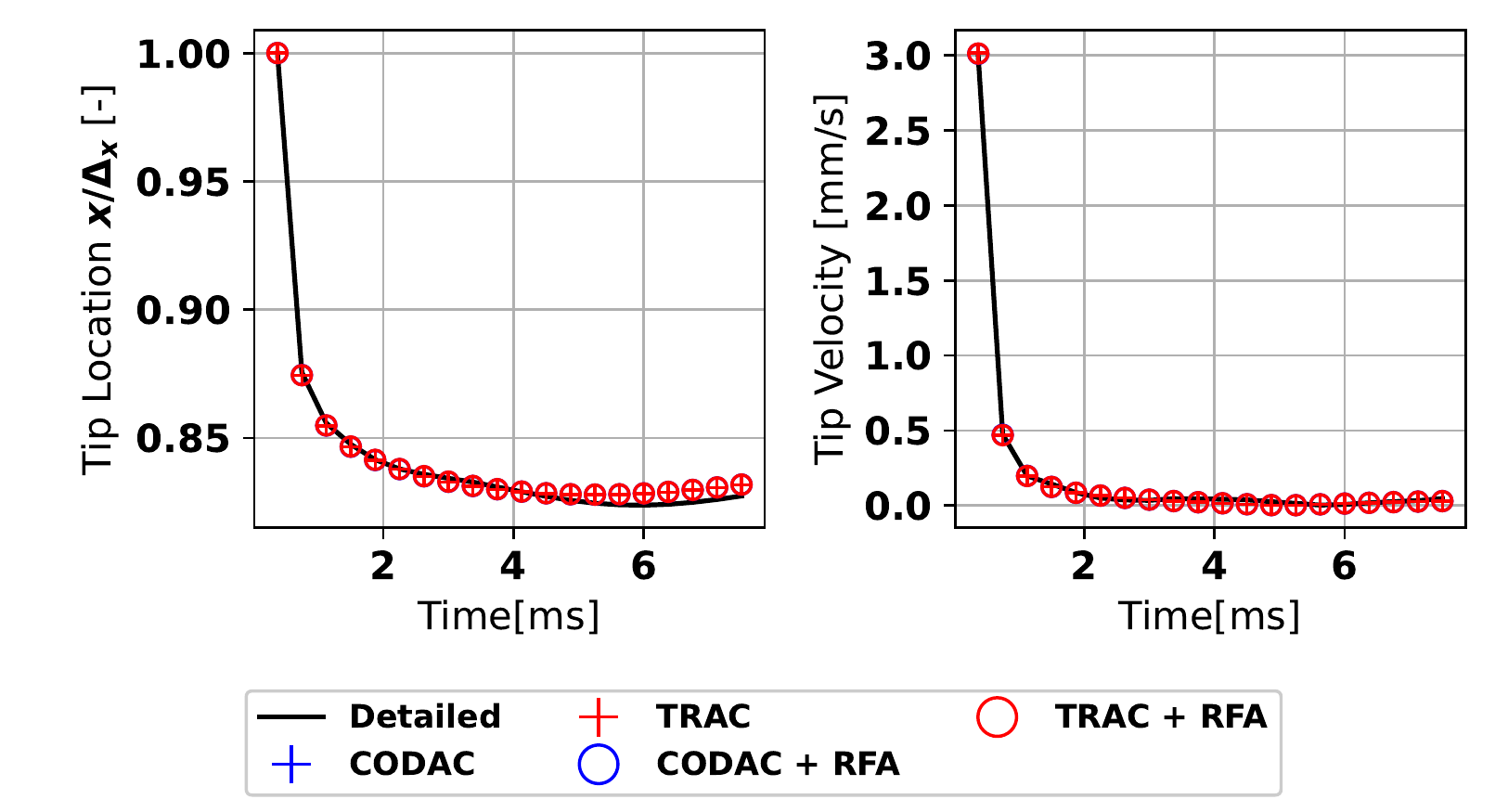}
   \vspace{8 pt}
 \caption{Temporal evolution of flame tip location and velocity in the triple flame for detailed and DAC methods.}
 \label{fig:XVEL}
\end{figure}

The regimes of the triple flame can be identified by the distribution of the source term of the progress variable ($\dot{\omega}_{Y_{c}}$) along mixture fraction and scaled-progress variable (C) iso-lines, as shown in Fig.~(\ref{fig:tripleReeess}). The scaling of the progress variable is obtained as $Y_{C}/Y_{{C}}^{max}$, where $Y_{{C}}^{max}$ corresponds to the local maximum value of the progress variable at the current instant. Iso-lines at $Y_{\xi}$ corresponding to equivalence ratios of $\phi$ = 1 and $\phi$ = 0.7 represent premixed and partially premixed burning, respectively. The iso-lines of C span across the entire flame capturing all regimes of combustion. In particular, the iso-line at C = 0.7 is dominated by flame propagation as it occurs in premixed combustion, while at C = 0.5, a trailing diffusion flame is generated from the interaction between the oxidizer and the combustion products of the rich premixed front.
The results of the TRAC method show good agreement when compared with CODAC and detailed chemistry, as seen in Fig.~(\ref{fig:tripleReeess}). The results indicate that the error introduced by the chemistry reduction and the mechanism tabulation approach has a low impact on the flame propagation. This confirms that the TRAC tabulation strategy is suited to capture accurately the reaction rates at all combustion regimes during transient conditions. The temporal evolution of the triple flame, represented by the leading edge velocity and location are shown in Fig.~(\ref{fig:XVEL}).  The flame tip is defined at the farthest upstream location at which a progress variable ($Y_{C} \, = \, 0.8 \, Y_{C,Prem}$) is found, where $Y_{C,Prem}$ is the maximum value of the progress variable obtained in a freely propagating stoichiometric laminar flame at similar conditions. The flame initially propagates into the mixture and it encounters the incoming velocity from the left boundary. Rich and lean flame fronts are formed on the leading edge at both sides of the stoichiometric point.

The transient evolution of the flame tip location and velocity illustrated by Fig.~(\ref{fig:XVEL}) shows that TRAC can capture accurately transient conditions with variations in flame propagation and heat release. The results demonstrate the robustness and accuracy of the TRAC method in multiregime combustion problems.

\begin{figure}[H]
\centering
  \includegraphics[width=0.65\linewidth]{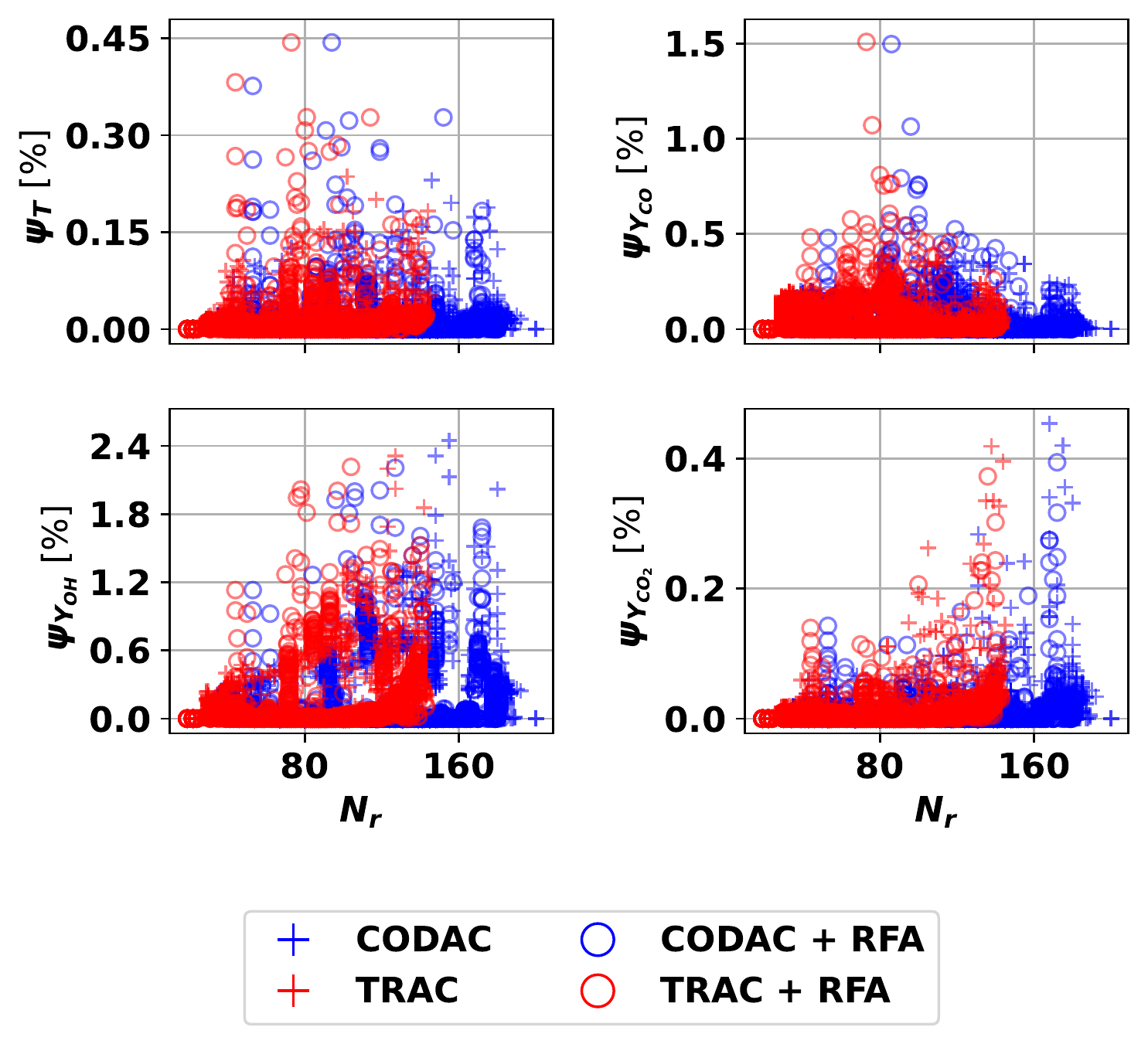}
    \vspace{6 pt}
  \caption{Error in $\%$ against number of active reactions $N_{r}$. Temperature (top left), mass fraction - $\mathrm{{CO}}$ (top right), mass fraction - $\mathrm{{OH}}$ (bottom left), and mass fraction - $\mathrm{CO_{2}}$ (bottom right).}
  \label{fig:errorNTriple}
\end{figure}

The error introduced by the chemistry reduction is, in general, a consequence of the eliminated reaction paths. A local measure of the error can be defined by comparing the instantaneous solutions from TRAC and CODAC with the reference solution. The number of reactions and their associated errors is shown in Fig.~(\ref{fig:errorNTriple}) for different scalar fields like temperature (T) and mass fractions of certain species, namely, $\mathrm{CO}$, $\mathrm{OH}$ and $\mathrm{CO_{2}}$. The data was extracted at a time of $2.5$ ms, which corresponds to the time instant at which the vortices are completely dispersed and no longer influence the mixing field. Furthermore, the selection of a time instance towards the end of the simulation allowed to account for the maximum possible accumulated error in the state variables (T, $Y_{k}$).

The TRAC and CODAC methods introduced less than 3 $\%$ of an error on average, with the maximum error seen for $\mathrm{OH}$. Despite this error, results show that the most important features of the flame, like burning velocity, flame structure, and heat release rates are accurately recovered. Species with the highest error in concentrations using the chemistry reduction methods were $\mathrm{HCNO}$, $\mathrm{HNCO}$, $\mathrm{NO}$, $\mathrm{H}$, $\mathrm{O}$ and $\mathrm{HO_{2}}$, which correspond to species related to $\mathrm{NO_{X}}$ oxidation pathways and radicals. Major species were predicted accurately and this was achieved by employing around 1/3 the number of reactions used by the reference mechanism, and make TRAC and CODAC methods very attractive for practical applications. The computational performance of these methods is discussed in the subsequent section.

\subsection{Computational Analysis}
\begin{figure}[H]
\centering
\begin{subfigure}{1.0\textwidth}
  \centering
  \includegraphics[width=0.49\linewidth]{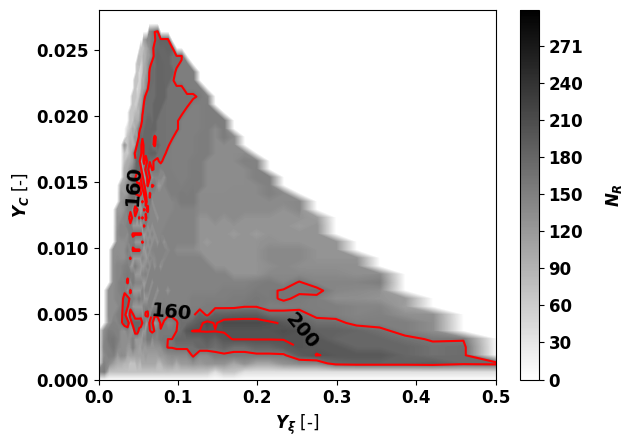}
  \label{fig:tripletabNorm}
  \includegraphics[width=0.49\linewidth]{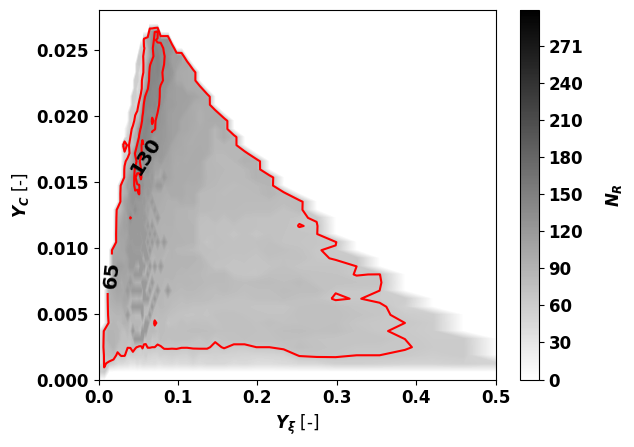}
  \label{fig:tripletabrfa}
\end{subfigure}
  \vspace{8 pt}
\caption{Visualization of the TRAC table with unscaled progress variable $Y_{c}$ vs mixture fraction $Y_{\xi}$ contoured by number of reactions PFA (left) and PFA + RFA (right), respectively.}
\label{fig:tripletab}
\end{figure}

This section presents a computational analysis of the TRAC and CODAC methods using the results of the triple flame. A visual representation of the TRAC tables is shown in Fig.~(\ref{fig:tripletab}). It shows the number of reactions contained in the reduced mechanisms stored in the database. It is seen that the tabulated reduced mechanisms of PFA + RFA are substantially smaller than those obtained by just the PFA method. A closer look at the PFA - TRAC table reveals two distinct clusters of reduced schemes with a high number of reactions: 1) near the stoichiometric point and 2) for rich mixtures at low temperatures. This is a consequence of the local normalization. The reduction is normalized locally to the most reactive pathway, irrespective of its global contribution. This correction increases the robustness of the method and its error control but may lead to regions in the database with reactions  with lower contributions to the chemical rates that could otherwise be removed. Reduction for such states is performed by RFA. The dependency of RFA on the production and consumption of key species shows that the reduction is more severe in states with lower contributions to the chemical rates. States with high chemical reactivity often result in more detailed reaction mechanisms.

\begin{figure}
\centering
  \includegraphics[width=0.49\linewidth]{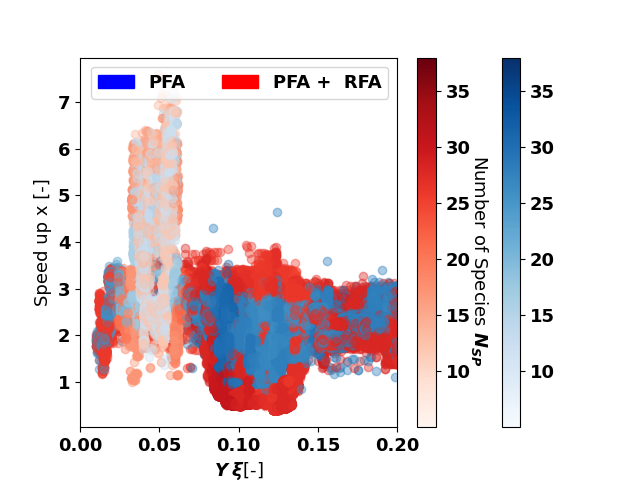}
  \includegraphics[width=0.49\linewidth]{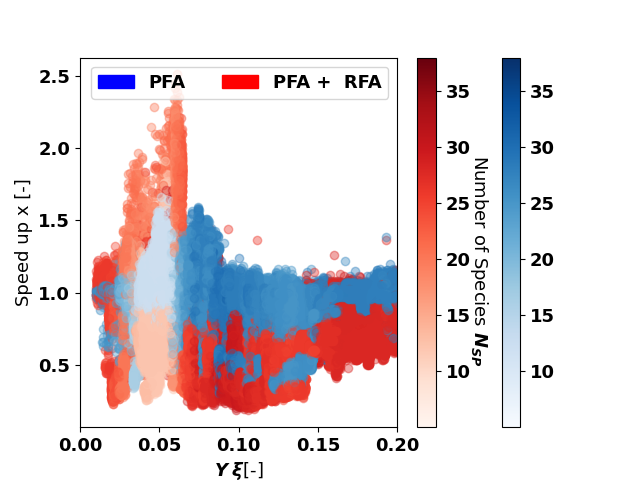}  
  \includegraphics[width=0.49\linewidth]{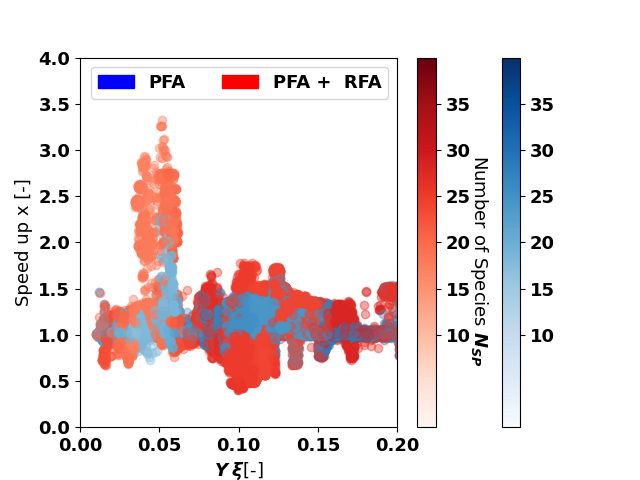}
  \includegraphics[width=0.49\linewidth]{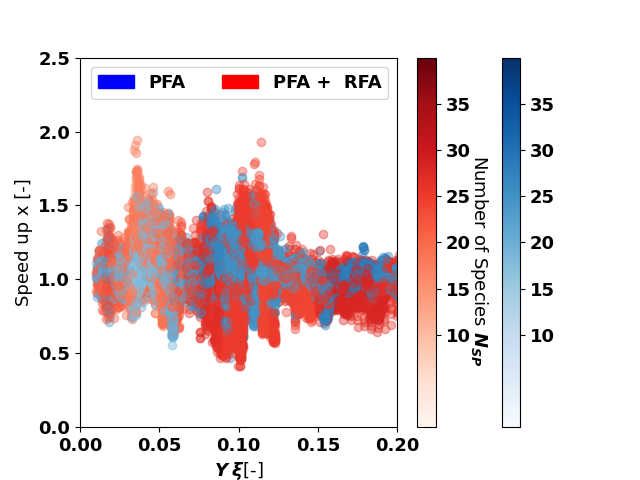}
    \vspace{6 pt}
  \caption{Speedup conditioned to  mixture fraction $Y_{\xi}$ in the triple flame. ODEPIM - TRAC (top left), CVODE - TRAC (top right), ODEPIM - CODAC (bottom left), and CVODE - CODAC (bottom right). Points are colored red (PFA + RFA) and blue (PFA) scaled to the number of active species.}
 
  \label{fig:cpuNrea2}
\end{figure}

The cost of chemical integration is a function of the chemical state and the number of reactions. Fig.~(\ref{fig:cpuNrea2}) shows the computational time of the chemical integration against $Y_{\xi}$ for the CODAC and TRAC methods, normalized to the computational time of the detailed chemistry solution. For both methods, reduction based on PFA and PFA + RFA is presented, which allows us to quantify the speedup achieved by the additional reduction using RFA. 

The computational performance was assessed using two different chemistry solvers: a variable order fully implicit scheme - CVODE \cite{cohen1996cvode} and a semi-implicit scheme - ODEPIM \cite{yang2017parallel}. This choice was motivated by numerous works \cite{lu2005directed,gao2015dynamic,savard2015computationally} published in the literature, which exposes the limitations of using a fully implicit solver to achieve additional 
performance by chemistry reduction, so a semi-implicit method is added to evaluate the effect.

\begin{table}[H]
\centering
\begin{tabular}{|l|l|l|l|l|}
\cline{1-4}
 Solvers  & Speedup   & nje & nst \\ \cline{1-4}
 CVODE  & 1.0 & $\approx$ 3 &  $\approx$ 5   \\ \cline{1-4}
 CVODE - TRAC & 1.21 & $\approx$ 4  &  $\approx$ 4  \\ \cline{1-4}
 CVODE - CODAC & 1.17 & $\approx$ 4  &  $\approx$ 4  \\ \cline{1-4}
 ODEPIM  & 2.4 & - & $\approx$ 3   \\ \cline{1-4}
 ODEPIM - TRAC & 4.56 & - & $\approx$ 3    \\ \cline{1-4}
 ODEPIM - CODAC & 1.24 & - & $\approx$ 3    \\ \cline{1-4}
\end{tabular}
\caption{Computational statistics, speedup w.r.t to CVODE solver, nje: number of jacobian evaluations and nst: number of solver iterations.}
\label{Tab:normalisedCPU}
\end{table}

The results indicate that the highest degree of reduction is found in states closer to stoichiometry regardless of the type of solver and DAC method, see Fig.~(\ref{fig:cpuNrea2}). The reactivity of the mixture is highest close to stoichiometry, 
where most of the computational time is spent in evaluating the minor species and radicals, which may not contribute significantly to the global burning rates. Reduction methods like PFA and RFA eliminate such reactions and only retain reaction pathways of significance. 

Speedups in the CVODE solver come as a consequence of the reduced number of reaction rate evaluations and ease of computing the jacobian for the non-linear solver. This is due to the fact that the PFA method does not reduce the stiffness of the chemistry and CVODE always assumes a dense jacobian with a dense solver. Moreover, this speedup is highly dependent on the thermo-chemical state and the selected reduced scheme. As seen in Fig.~(\ref{fig:cpuNrea2}), there are points in which the computational cost of integrating the reduced schemes is higher than those of the detailed scheme, which is due to the local increase of stiffness generated by the reduction process in some locations. But in general, there is a substantial gain in the overall performance. The results also show higher speedups for the semi-implicit solver (ODEPIM), as it does not involve the solution of a non-linear system. For semi-implicit schemes such as ODEPIM, the cost scales with $\mathcal{O}({N_{sp}} \, c)$, where $N_{sp}$ is the number of species and $c$ takes values between $1$ and $4$, based on the number of iterations the solver takes. Similar trends are also observed in Fig.~(\ref{fig:cpuNrea2}). It shows that states away from stoichiometry are represented by reduced schemes either of the same size or even larger than those used near stoichiometry. This is a consequence of the local normalization of the relation matrix for PFA, which makes the reduction always relative to the local state. Reduction based on global states might help achieve higher speedups in these states. However, since the overall cost of integration in these states is fractional to the cost near stoichiometry, the overall gain would be marginal. Moreover, the normalization adds robustness to the definition of error in the reduction and this can be preferred over minimal gains in speedup. 

Additional reduction in chemistry  using RFA with CVODE shows higher speedups. This is attributed to the speedup of computation of the jacobian for CVODE and the solutions of the non-linear system. For the ODEPIM solver, the additional speedup with RFA is marginal, since 
it scales with the number of species $N_{sp}$ and RFA only reduces the number of reactions. The speedups (w.r.t to the detailed CVODE solver), number of jacobian evaluations (nje), and solver iterations (nst) are summarized in Tab.~(\ref{Tab:normalisedCPU}) for all the relevant cases. While the number of jacobian evaluations using TRAC with CVODE increases, this additional cost is balanced by the reduced cost of computing the jacobian and fewer iterations of the solver. Despite the similar number of sub-iterations of the ODEPIM solver for both the PFA and PFA + RFA solutions, there is an additional speedup in the PFA + RFA solution due to the lower number of active reactions.   

The cost of the CODAC method also includes the cost of reduction, thus the peak speedup achieved by the method is a factor 3x, which is lower than that of TRAC with a factor 7x when solved with ODEPIM. Note that the difference between the methods is smaller with CVODE. 
This difference can be addressed by taking a look at the ratio of CPU costs ($\kappa = \tau_{Red} / \tau_{Int}$), where $\tau_{Red} $ and $\tau_{Int}$ are the CPU cost of reduction and chemistry integration, respectively. The value of $\kappa_{ODEPIM}$ is about 9.3, while $\kappa_{CVODE}$ is around 2.1. This indicates that the difference in speedups between the solvers comes fundamentally from the chemical integration. Hence, it is essential to use schemes that are sensitive to the reduced order system 
to maximize the speedups achieved by the chemistry reduction.

\section{Conclusions}\label{Sec:4}\addvspace{10pt}

A novel Dynamic Adaptive Chemistry (DAC) method based on the tabulation of locally reduced reaction mechanisms (TRAC) is presented in this study. The proposed strategy uses the concept of a low-dimensional database to define a chemical space to store reduced chemical schemes as a function of a given set of control variables. In this study, the mixture fraction $Y_{\xi}$ and a chemical progress variable $Y_{C}$ are used to associate the chemical states with their corresponding local reduced reaction mechanism. Chemistry reduction is achieved by a combination of PFA with RFA, which can be performed either during runtime or in a pre-processing step. A logarithmic relation is proposed to better  correlate the PFA threshold with the reduction error. This motivated the definition of a more universal expression based on a normalization process. RFA was found to work well with PFA for additional reduction, without introducing significant errors. The novel TRAC method is compared with  CODAC and detailed chemistry in canonical transient cases. A homogeneous auto-ignition case was used to validate the method and evaluate the influence of the degree of reduction with an \emph{a priori} evaluation. The study was then extended to extinguishing and auto-igniting counter flow diffusion flames, in which the transient performance of the TRAC method is examined.
The results show a good correlation between major species and temperature, even $\mathrm{CO}$ was correctly predicted for TRAC with PFA. The performance and accuracy of TRAC are addressed in the transient problem of a triple flame. Instantaneous and transient characteristics of the flame were accurately recovered. The error introduced by the reduction is plotted against the number of reactions in both TRAC and CODAC with PFA and PFA + RFA reduction strategies respectively. The maximum error was found to be around 3$\%$, and it was in the prediction of $\mathrm{OH}$. A speedup with a factor of 4x was achieved with the TRAC method when used in combination with a semi-implicit solver. Results obtained with the novel TRAC method show promise for use in detailed reacting flow simulations of transient thermo-chemical problems. TRAC does not require prior knowledge of the relevant species and reactions to define appropriate error functions to store the reduced schemes and it adapts well to transient and multiregime problems. Although, fine-tuning the thresholds and identification of relevant species in simplified versions of the complex problem of interest will substantially improve the accuracy of  the method. TRAC can also be applied to target prediction of specific phenomena like pollutants, however this would require not only the exploration of the impact of higher-order relations for PFA, but also dimensions of the tabulation space of the reaction mechanisms and the possibility of more controlling variables. This aspect is left out of the scope of the present work, and will be the focus of future studies. Lastly, the use of
efficient High-Performance Computing (HPC) algorithms for chemical integration, chemistry reduction, and transport is vital to make TRAC a feasible option to simulate complex reacting flows.

\section{Acknowledgments}\label{Acknowledgments}\addvspace{10pt}

This project has received funding from the European Research Council (ERC) under the European Union’s Horizon 2020 research and innovation program (grant agreement No. 682383), the Center of Excellence in Combustion project (grant agreement No 952181), and the AHEAD PID2020-118387RB-C33 and ORION TRA2017-89139-C2-2-R projects from the Ministerio de Ciencia e Innovaci\'on. Anurag Surapaneni acknowledges the support from the Secretariat for Universities and Research of the Ministry of Business and Knowledge of the Government of Catalonia and the European Social Fund.

\label{}

\label{}

\section{References}
\bibspacing=\dimen 100
\bibliographystyle{ieeetr}
\bibliography{mybibfile}

\end{document}